\documentclass[sigconf]{acmart}
\usepackage{amsmath}
\usepackage{natbib}
\usepackage{float}
\usepackage{graphicx}
\usepackage{epstopdf}
\usepackage{subfigure}
\usepackage{amsthm}
\usepackage{enumerate}
\usepackage{multirow}
\usepackage[ruled,linesnumbered]{algorithm2e}  
\usepackage{algorithmic}
\usepackage{bbm}
\usepackage{bm}
\newtheorem{theorem}{Theorem}

\newtheorem{proposition}{Proposition}

\usepackage{balance}

\AtBeginDocument{%
  \providecommand\BibTeX{{%
    \normalfont B\kern-0.5em{\scshape i\kern-0.25em b}\kern-0.8em\TeX}}}

\copyrightyear{2023} 
\acmYear{2023} 
\setcopyright{acmlicensed}\acmConference[RecSys '23]{Seventeenth ACM Conference on Recommender Systems}{September 18--22, 2023}{Singapore, Singapore}
\acmBooktitle{Seventeenth ACM Conference on Recommender Systems (RecSys '23), September 18--22, 2023, Singapore, Singapore}
\acmPrice{15.00}
\acmDOI{10.1145/3604915.3608797}
\acmISBN{979-8-4007-0241-9/23/09}

\begin{document}

\title{Uncovering User Interest from Biased and Noised Watch Time \\ in Video Recommendation}
% \affiliation{%

\author{Haiyuan Zhao}
\affiliation{
\institution{School of Information, Renmin University of China}
\city{}\country{}
}
\email{haiyuanzhao@ruc.edu.cn}

\author{Lei Zhang}
\affiliation{%
  \institution{Gaoling School of Artificial Intelligence, Renmin University of China}
\city{}\country{}
  }
\email{zhanglei1010@ruc.edu.cn}

\author{Jun Xu}
\authornote{Jun Xu is the corresponding author. Work partially done at Engineering Research Center
of Next-Generation Intelligent Search and Recommendation, Ministry of Education.
}
\affiliation{%
  \institution{Gaoling School of Artificial Intelligence, Renmin University of China}
\city{}\country{}
  }
\email{junxu@ruc.edu.cn}

\author{Guohao Cai}
\affiliation{%
  \institution{Noah’s Ark Lab, Huawei}
\city{}\country{}  }
\email{caiguohao1@huawei.com}

\author{Zhenhua Dong}
\affiliation{%
  \institution{Noah’s Ark Lab, Huawei}
\city{}\country{}
  }
\email{dongzhenhua@huawei.com}

\author{Ji-Rong Wen}
\affiliation{%
  \institution{Gaoling School of Artificial Intelligence, Renmin University of China}
\city{}\country{}
  }
\email{jrwen@ruc.edu.cn}

\renewcommand{\shortauthors}{Haiyuan Zhao et al.}


\begin{abstract}
In the video recommendation, watch time is commonly adopted as an indicator of user interest. However, watch time is not only influenced by the matching of users' interests but also by other factors, such as \textbf{duration bias} and \textbf{noisy watching}. Duration bias refers to the tendency for users to spend more time on videos with longer durations, regardless of their actual interest level. Noisy watching, on the other hand, describes users taking time to determine whether they like a video or not, which can result in users spending time watching videos they do not like.
Consequently, the existence of duration bias and noisy watching make watch time an inadequate label for indicating user interest. Furthermore, current methods primarily address duration bias and ignore the impact of noisy watching, which may limit their effectiveness in uncovering user interest from watch time.
In this study, we first analyze the generation mechanism of users' watch time from a unified causal viewpoint. Specifically, we considered the watch time as a mixture of the user's actual interest level, the duration-biased watch time, and the noisy watch time. To mitigate both the duration bias and noisy watching, we propose \textbf{D}ebiased and \textbf{D}enoised watch time \textbf{Co}rrection (D$^2$Co), which can be divided into two steps: First, we employ a duration-wise Gaussian Mixture Model plus frequency-weighted moving average for estimating the bias and noise terms; then we utilize a sensitivity-controlled correction function to separate the user interest from the watch time, which is robust to the estimation error of bias and noise terms. The experiments on two public video recommendation datasets and online A/B testing indicate the effectiveness of the proposed method. 
% We also theoretically and empirically proved that our method outperforms other baseline models by mitigating both duration bias and duration noise.
\end{abstract}



\maketitle

\section{Introduction}

% \begin{enumerate}[(1)]
% \item Why use watch time as the substitute of user interest in video recommendation
% \item The problems of watch time: duration bias and duration noise
% \item Current method for correcting watch time, their mechanism and drawbacks
% \item Our analysis about watch time and Our proposed method: XXX.
% \item Contributions of our study
% \end{enumerate}

The rising of video content platforms has attracted billions of users and become more frequent in the daily use of users nowadays~\citep{Covington2016Deep,Davidson2010YouTube,Liu2019User,Liu2021Concept}. In order to better satisfy the information needs of users and improve their engagement, an accurate and personalized video recommendation plays a significant role. Unlike the traditional recommendation scenario, the video recommendation adopts a streaming play pattern~\citep{gao2022kuairand,Gong2022Real}. That is, a recommender system switches to the next video and plays it automatically when the user finishes playing the previous one. This feature makes the widely used implicit feedback (e.g., user click) no longer suitable as a label to measure user interest. Compared to clicks, users' watch time indicates how much attention the user is willing to spend on this video and has been considered a better indicator of user interest~\citep{Covington2016Deep,wu2018beyond,wang2020capturing,Yi2014Beyond}.

% \textcolor{blue}{User Generated Content platforms such as TikTok and Youtube have attracted billions of consumers, surfing in the plentiful short-videos every day.}

\begin{figure}
    \subfigure[Duration Bias]{
    \includegraphics[width=0.22\textwidth]{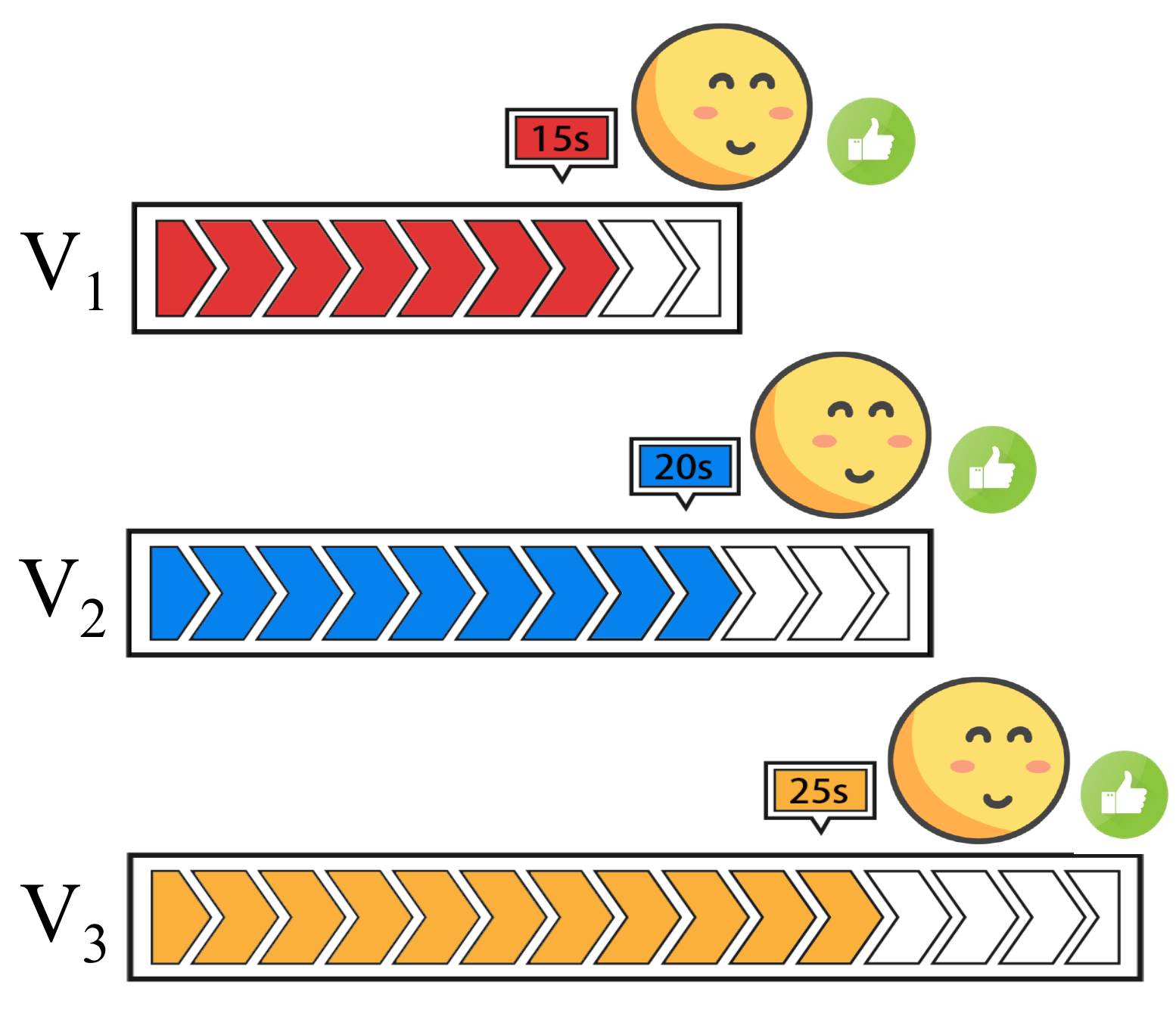}}
    \quad
    \subfigure[Noisy Watching]{
    \includegraphics[width=0.22\textwidth]{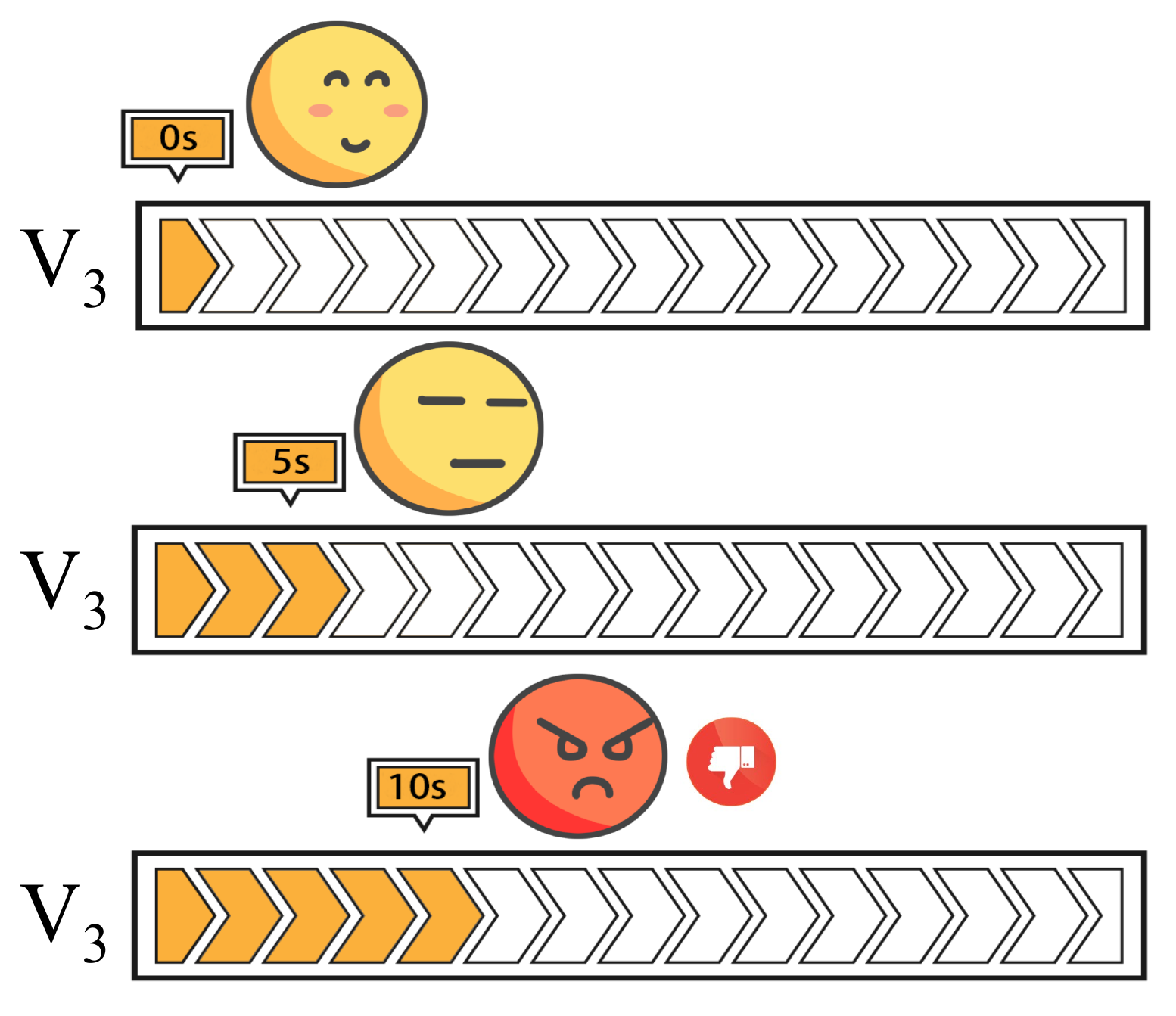}}
    \caption{The illustration of duration bias and noisy watching. (a) the user watches different videos. (b) the user watches the same video.}
    \label{fig: illustration of bias and noise}
\end{figure}

However, the length of watch time is not only determined by user interest alone but also affected by other non-interest factors. 
On the one hand, users tend to spend more time watching engaging videos with longer durations, resulting in longer average watch time for long videos. This phenomenon is referred to as \textbf{duration bias}~\citep{Zhan2022Deconfounding, Zheng2022DVR}. As shown in Fig.~\ref{fig: illustration of bias and noise}(a), all three videos $v_1, v_2,v_3$ are of interest to users but have different durations. It can be seen that users have a longer watch time for engaging videos with longer duration (e.g., $v_3$). If we regard watch time as the indicator of user interest, the duration bias will mislead the recommendation models leans to recommend more long videos. On the other hand, users need time to perceive whether they like newly recommended videos. As a result, they may watch videos they are not interested in for a while, commonly referred to as \textbf{noisy watching}~\citep{Li2022Improving}. Fundamentally, noisy watching results from the users' trust in the recommender system itself~\citep{Agarwal2019Addressing} or the clickbait content at the beginning of videos~\citep{Wang2021Clicks}. As shown in Fig.~\ref{fig: illustration of bias and noise}(b), users tend to believe that the newly recommended videos engage them when the video starts playing. Consequently, they may begin watching this video and take some time (e.g., 10s) to realize they are not interested in it. The presence of noisy watching results in users spending time watching videos they do not like, which can also mislead the recommendation models if we regard watch time as the indicator of user interest.

\begin{figure}
    \subfigure[Mean watch time for videos of interest]{
    \includegraphics[width=0.22\textwidth]{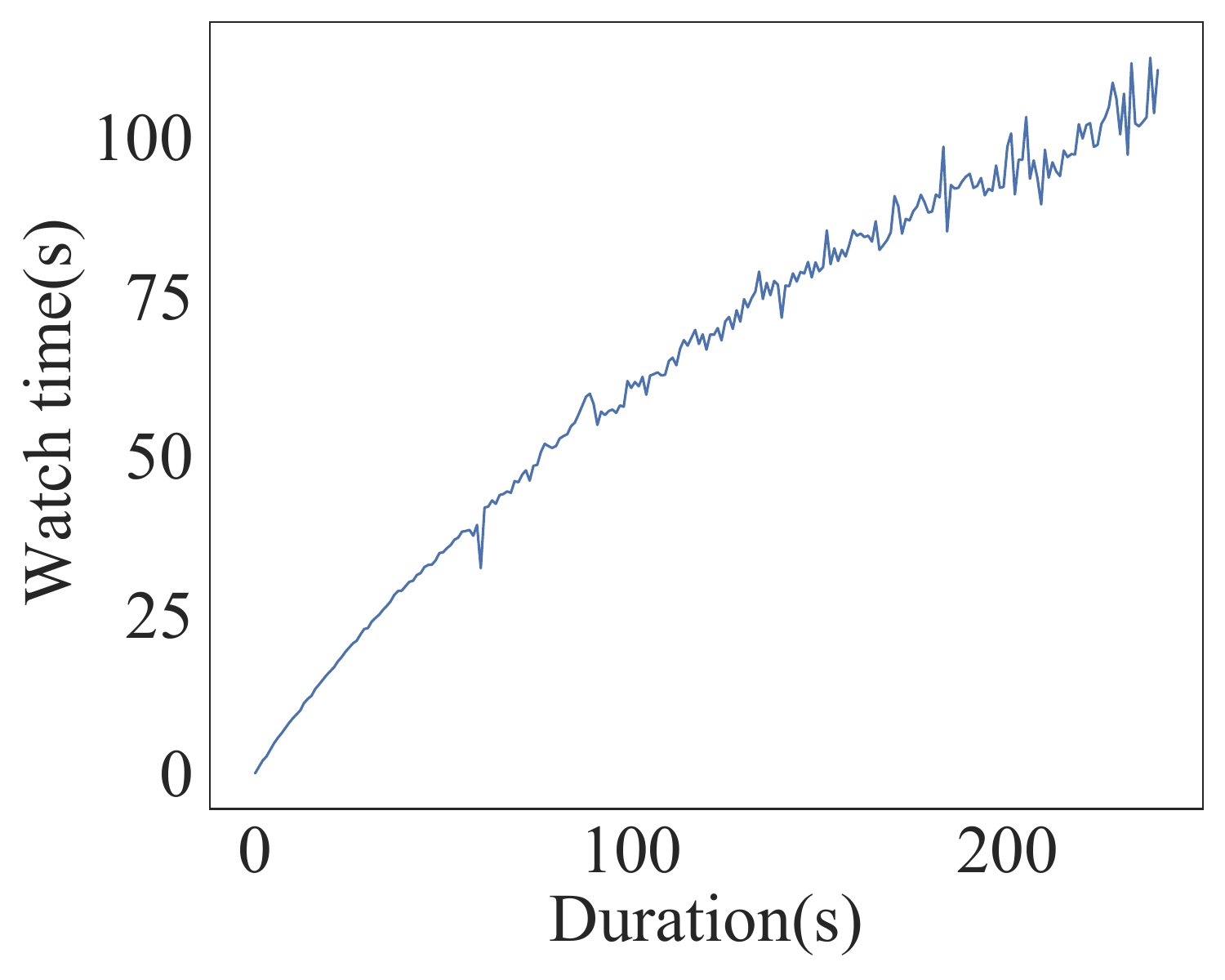}}
    \quad
    % \hspace{.1in}
    \subfigure[Mean watch time for videos not of interest]{
    \includegraphics[width=0.22\textwidth]{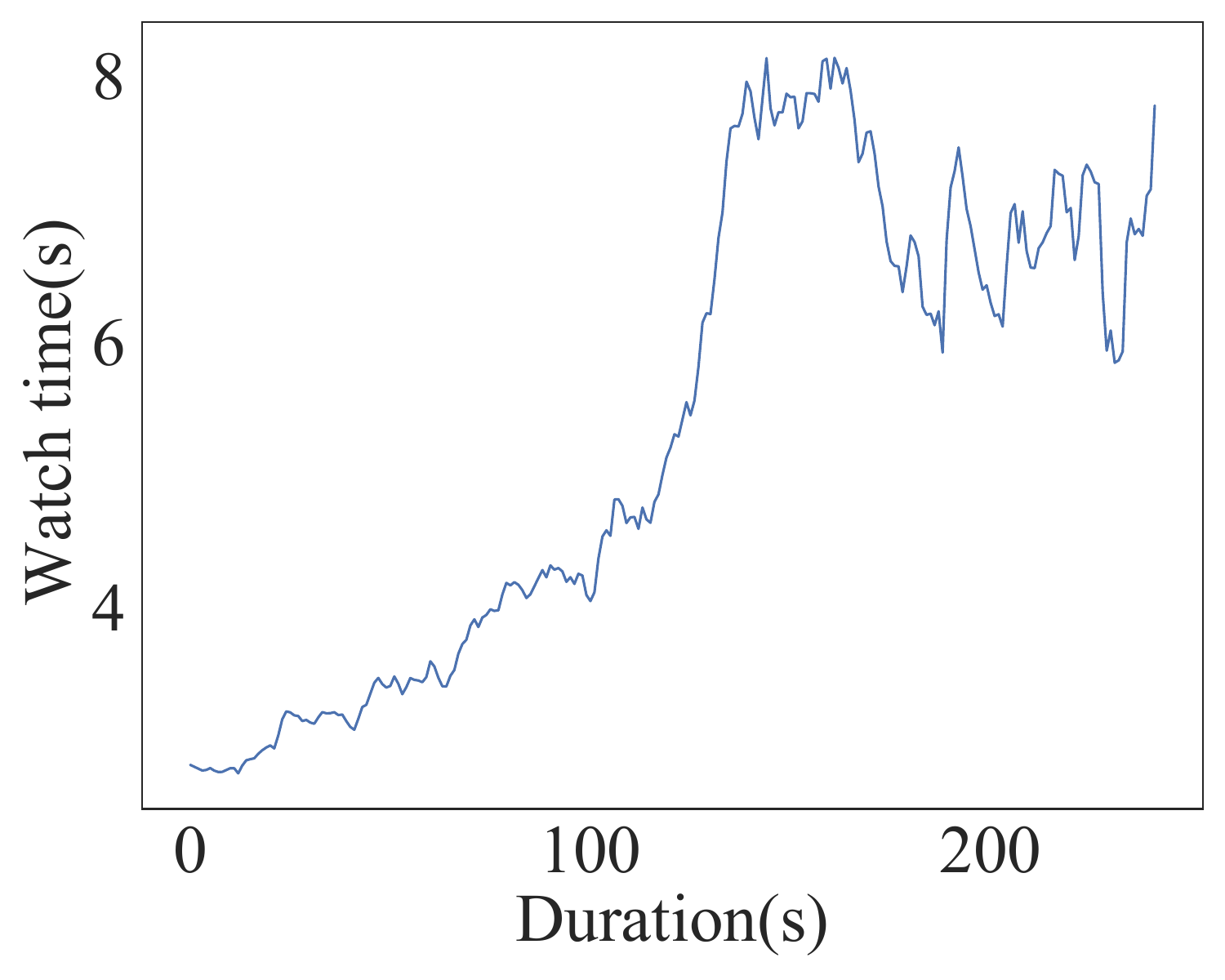}}
    \caption{The evidence of the existence of duration bias and noisy watching in the subset of the KuaiRand dataset. We calculate the mean watch time for videos that are/aren't interesting to users in different duration.}
    \label{fig: evidence of bias and noise}
\end{figure}

To verify the existence of the aforementioned duration bias and noisy watching, we conducted a pilot study on the KuaiRand dataset~\citep{gao2022kuairand}, a large-scale public video recommendation dataset collected from Kuaishou. For detecting the duration bias, we first aim to find records in the dataset that are engaging to users. Although we do not know users' latent interest behind each record, there is still some behavior feedback~\citep{Zhao2019Recommending} in the dataset. Specifically, we treat one record as of interest to the user if one of the positive behavior feedback in \emph{like}, \emph{follow}, \emph{forward}, \emph{comment}, \emph{profile enter} is presented. Then, we calculate the mean watch time in different duration on this subset. As shown in Fig.~\ref{fig: evidence of bias and noise}(a), the mean watch time of engaging videos increases with the duration growth, which verified the existence of duration bias. Similarly, for detecting noisy watching, we first regard records with negative behavior feedback like \emph{hate} as not engaging to the user. Then, we calculate the mean watch time on this subset. As shown in Fig.~\ref{fig: evidence of bias and noise}(b), the mean watch time of videos that users aren't interested in is not zero, verifying the existence of noisy watching. Meanwhile, we can find that the curve in Fig.~\ref{fig: evidence of bias and noise}(b) increases as the duration grows. This is because longer videos usually have richer video content or a more prolonged beginning, which makes users spend more time perceiving their level of interest.

Despite the hazards, duration bias and noisy watching are much less explored as compared to many other biases in recommender system research. One heuristic way to address duration bias is to divide the watch time by the video duration, called Play Complete Rate (PCR). However, it is worth noting that the trend between watch time and duration is not a simple linear relationship according to Fig.~\ref{fig: evidence of bias and noise}. Therefore, simply dividing by duration cannot eliminate the duration bias.
To better address duration bias, \citet{Zhan2022Deconfounding} proposed to transform normal watch time prediction into duration-grouped watch time quantile to mitigate the negative effects of duration bias. \citet{Zheng2022DVR} proposed standardizing the watch time according to different video duration and leveraging the standardized score as the supervision signal to train and evaluate the video recommendation model. Although effective, there still has much space for improvement: (i) current studies only focus on addressing the duration bias while overlooking the noisy watching, which makes their predicted user interest signals still inaccurate; (ii) Existing approaches (e.g., \cite{Zhan2022Deconfounding} and~\cite{Zheng2022DVR}) rely on underlying assumptions (we will discuss this in section~\ref{subsec: baseline analysis}) about the distribution of user interests for correcting the duration bias. Once these assumptions are violated in practice, their performances cannot be guaranteed. % a containing proposed approach still far from truly separating user interest from watch time.

To jointly model both duration bias and noisy watching, we first conduct a causal analysis of the generation mechanism of users' watch time. Unlike current methods, which only notice the duration bias in watch time, we considered the watch time as a mixture of the user's actual interest level, the duration-biased watch time, and the noisy watch time. Then we propose a model called \textbf{D}ebiased and \textbf{D}enoised watch time \textbf{Co}rrection (D$^2$Co) to mitigate the duration bias and noisy watching. Specifically, we propose to regard the distribution of watch time in each duration length as a mixture of latent bias and noise distributions. A duration-wise Gaussian mixture model is employed to estimate the parameters of these latent distributions. Since the adjacent value of duration should have similar properties, a frequency-weighted moving average is used to smooth the estimated bias and noise parameters sequence. Then we utilized a sensitivity-controlled correction function to separate the user interest from the watch time, which is robust to the estimation error of bias and noise parameters.
% Finally, we utilize a sensitivity-controlled correction function, \xujun{instead of a standard affine correction} 
%这句话显得没头没脑的，突然冒出来}
% , to better separate user interest from watch time.% for further enhancing our method. 

Compared to existing methods, D$^2$Co enjoys the advantages of correcting the duration bias and noisy watching simultaneously in video recommendation and does not require critical assumptions on the distributions of the user interest. %We verified its effectiveness both theoretically and empirically. 
The major contributions of the paper include the following:
\begin{enumerate}[(1)]
\item We analyze the existence of duration bias and noisy watching in the video recommendation and provide a unified causal view for modeling the bias and noise simultaneously.
\item We propose D$^2$Co, a method for mitigating both the duration bias and noisy watching. D$^2$Co can obtain user interest from watch time and does not rely on the critical assumption of user interest distribution.
\item We conducted offline experiments on two public video recommendation datasets and an online A/B test on the real video product. Experimental results verified the effectiveness of the proposed model and theoretical conclusions. 
\end{enumerate}

\section{Related Works}

% \subsection{Video Recommendation}
\textbf{Video Recommendation}~With the rapid growth of video content, personalized recommendation is widely used to provide videos of interest to users in video applications. The key challenge for video recommendation is to mine user interest from various signals~\citep{Wang2019Overview}. In a classic recommendation scenario, Click-Through-Rate (CTR) is an effective metric for measuring user interest~\citep{Rendle2012Factorization, Cheng2016Wide,guo2017deepfm,Song2019AutoInt}.
% users will actively select the items they are interested in and click on them. 
However, since the video recommendation scenario adopts a streaming play pattern, 
% That is, when users finish watching the previous video, the newly recommended video will automatically play. This difference 
clicks are no longer a reliable indicator of user interest. Instead, users' watch time is commonly used as a substitute indicator of user engagement~\citep{Covington2016Deep,wu2018beyond,wang2020capturing,Yi2014Beyond}. For instance, \citet{Covington2016Deep} treated the watch time as a weight of each impressed video and utilized a weighted logistic regression for predicting watch time. \citet{wu2018beyond} investigated the bias of watch time as well as watch percentage from an aggregated level and defined relative engagement to measure the video quality. Moreover, other trials utilized multiple user behaviors to enhance video recommendation. For example, \citet{Zhao2019Recommending} proposed a large-scale multi-objective ranking system for recommending what video to watch next on an industrial video-sharing platform. \citet{Li2019Routing} designed a graph-based sequential network to simultaneously model users' dynamic and diverse interests. 
% It is able to capture user interest by leveraging multiple user behaviors. 
\citet{Wei2019MMGCN} considered the interactions between users and items and the item contents from various modalities.
% , and proposed a multi-modal graph convolution network framework for capturing user preferences.

%\subsection{Debiasing in Information Retrieval}
\textbf{Debiasing in Information Retrieval}~Alleviating the bias is of great importance in current information retrieval systems. Most efforts are devoted to address the position bias~\citep{Joachims2017Unbiased,Ai2018Unbiased,Yuan2020Unbiased}, popularity bias~\citep{Zhang2021Causal,Zheng2021Disentangling,Wei2021Model} and selection bias~\citep{schnabel2016recommendations,Wang2016Learning,Saito2020Unbiased} in recent studies.  Inspired by causal inference~\citep{Wu2022survey}, a large number of debiasing methods are proposed for mitigating aforementioned biases, which includes propensity-based methods~\citep{Joachims2017Unbiased,Zhang2022Counteracting}, backdoor adjustment methods~\citep{Wei2021Model,Zhang2021Causal} and causal embedding methods~\citep{Zheng2021Disentangling,Chen2021Adapting,Bonner2018Causal}.  As we discussed before, the bias in video recommendation is mainly duration bias. However, only a few studies~\citep{Zhan2022Deconfounding,Zheng2022DVR} are focused on this bias in video recommendation. In contrast to our approach, existing methods for addressing duration bias rely on critical assumptions to achieve their unbiasedness.

% \subsection{Denoising in Information Retrieval }
\textbf{Denoising in Information Retrieval}
To denoise data for improving model performance has been an emerging research topic in recent years. In general, the noised data is defined as the false-positive and false-negative samples among the dataset. The core idea of current studies is to mine noise data based on \emph{Memorization Effect}~\citep{arpit2017closer}. That is, models can easily remember clean samples but have difficulty remembering noisy samples. For instance, \citet{Wang2021Denoising} tried to mine noisy samples from the loss value and designed an adaptive threshold mechanism for truncating these samples of high loss values. \citet{Wang2022Learning} proposed to discover noisy samples from the disagreement of different models. \citet{Gao2022Self} proposed a self-guided learning framework to collect memorized interactions at the early stage of the training. However, the above studies aim to develop a generic approach without specifically analyzing the noise in the video recommendation scenario.

\section{Problem Statements}

\subsection{Problem Formulization}

The problem of video recommendation can be described as follows. Given a user $u$ and a recalled video $v$ with duration $d$, each user-video pair $(u,v)$ is described by an $n$-dimensional feature vector $\mathbf{x}=\phi(u,v) \in\mathbb{R}^n$. The interest of $u$ in $v$ can be represented by an unobserved variable $R$. Without loss of generality, we assume that $R\in\{0, 1\}$ is a binary variable, which is sampled from latent Bernoulli distribution $\Pr(R=1\mid\mathbf{x})$. The users' watching behavior on videos can be recorded as the log data $\mathcal{D}=\{(\mathbf{x}_i, w_{i}, d_{i})\}_{i=1}^{N}$, where $\mathbf{x}_i, w_{i}, d_{i}$ respectively denote the $i$-th user-video pair's feature vector, user's watch time on this video, and the duration of this video (e.g., in seconds) 
% (\xujun{e.g., in seconds}). 

Ideally, a scoring function $f(\mathbf{x}):\mathbb{R}^n\rightarrow \mathbb{R}$ could be learned by minimizing the following ideal point-wise loss:
\begin{equation}
    \mathcal{L}_{\mathrm{ideal}} = \frac{1}{|\mathcal{D}|}\sum_{\mathcal{D}}-r\log\left[\sigma\left(f(\mathbf{x})\right)\right] - (1-r)\log\left[1-\sigma\left(f(\mathbf{x})\right)\right],
    \label{eq: ideal_loss}
\end{equation}
where $r$ is the unobserved user's true interest in a video, $\sigma$ is the sigmoid function. Equation~(\ref{eq: ideal_loss}) cannot be minimized because the interest indicator $r$ is unobserved. An alternative way is naively fitting the prediction to the observed watch time $w$ in $\mathcal{D}$:
\begin{eqnarray}
&&\!\!\!\!\!\!\!\!\!\!\!\!\!\!\!\!\!\!\!\!\!\!    \mathcal{L}_{\mathrm{naive}} =\frac{1}{|\mathcal{D}|}\sum_{\mathcal{D}} \!\!-\frac{w}{w_{\mathrm{max}}}\log\left[\sigma\left(f(\mathbf{x})\right)\right]\nonumber\\ 
&&- (1-\frac{w}{w_{\mathrm{max}}})\log\left[1-\sigma\left(f(\mathbf{x})\right)\right],
    \label{eq: naive_loss}
\end{eqnarray}
where $w_{\mathrm{max}}$ is the maximum watch time in the whole $\mathcal{D}$. Note that since the value of watch time $w$ is not between 0 and 1, it is scaled into the interval $[0,1]$ by simply dividing with $w_{\mathrm{max}}$. 
As has been discussed, there exists a gap between the optimal solution of $\mathcal{L}_{\mathrm{naive}}$ and that of $\mathcal{L}_{\mathrm{ideal}}$ because the watch time
$w$ suffers from both duration bias and noisy watching. The goal of this paper is to mitigate the bias and noise, i.e., uncovering the user interest from watch time for learning better scoring function $f(\mathbf{x})$.

\subsection{Causal Analysis of Watch Time}
\begin{figure}[t]
    \includegraphics[width=0.35\textwidth]{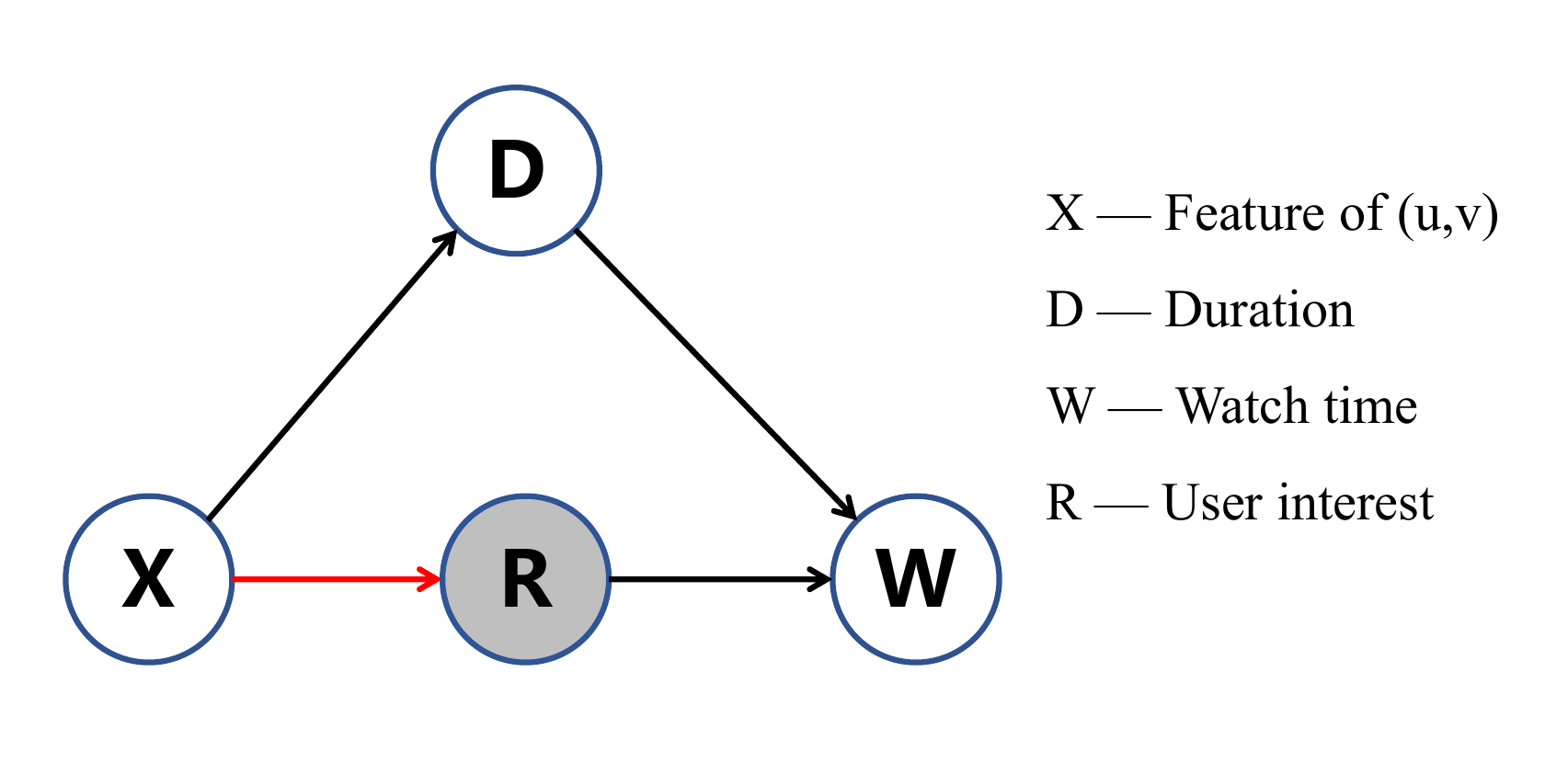}
    \caption{Causal graph of users' watch time in video recommendation. The gray node denotes the unobserved variable R. The red arrow denotes the effect that the recommendation model needs to estimate.}
    \label{fig:causal_graph}
\end{figure}

Next, we analyze how the duration bias and noisy watching affect the watch time based on the causal graph~\citep{pearl2009causality} shown in Figure~\ref{fig:causal_graph}. 
Given a user-video pair $(u,v)$, its feature vector $\mathbf{x}$ decides both duration $D$ and user interest $R$. This is reasonable because the video duration is part of the endogenous features of this video, and the level of user interest in this video can be considered as the matching extent between the user feature and the video feature. Then duration $D$ and user interest $R$ decide the watch time $W$ together, as we discussed before. Since the user interest $R$ is an unobserved variable in the dataset, watch time $W$ is leveraged as a surrogate label of $R$. Unfortunately, besides the relevance $R$, $W$ is also affected by the video duration $D$, which leads to duration bias and noisy watching. Therefore, directly fitting watch time $W$ will result in an erroneous video recommendation model.

According to this causal graph, we can formulate the expected watch time for a given user-video pair as follows:

\begin{equation}
\label{eq:causal_decompose}
\begin{aligned}
        \mathbb{E}(W\mid\mathbf{x})&=\sum_{W}w\Pr(W=w\mid \mathbf{x})
        \\&=\sum_{W}w\left(\!\!\sum_{D}\sum_{R\in \{0,1\}}\Pr(W=w\mid D,R)\Pr(D\mid\mathbf{x})\Pr(R\mid\mathbf{x})\!\!\right) \\
        &=\sum_{W}w\left(\sum_{R\in \{0,1\}}\Pr(W=w\mid d, R)\Pr(R\mid\mathbf{x})\right)
        \\&=\sum_{R\in \{0,1\}}\left(\sum_{W}w\Pr(W=w\mid d, R)\right)\Pr(R\mid\mathbf{x}) \\
        &=\sum_{R\in \{0,1\}}\mathbb{E}(W\mid d, R)\Pr(R\mid\mathbf{x}).
\end{aligned}
\end{equation}
The first equation is the definition of expectation; the second equation is the decomposition of $\Pr(W=w\mid \mathbf{x})$ based on the Figure~\ref{fig:causal_graph}; the third equation is based on the fact that one video only has a unique duration and the fourth equation is the multiplication switching law. Finally, we decomposed $\mathbb{E}(W\mid\mathbf{x})$ into the mixture of $\mathbb{E}(W\mid d, R=1)$ and $\mathbb{E}(W\mid d, R=0)$, which is weighted by $\Pr(R=1\mid\mathbf{x})$ and $\Pr(R=0\mid\mathbf{x})$, respectively. 

Specifically, the $\mathbb{E}(W\mid d, R=1)$ represents the average time users will watch a video of duration $d$ due to their interest, which indicates the length of duration-biased watch time. Meanwhile, the $\mathbb{E}(W\mid d, R=0)$ represents the average time users will watch a video of duration $d$ they are not interested in, which indicates the length of noisy watch time. $\Pr(R=1\mid\mathbf{x})$ indicates the user's interest level for a video. For the ease of notation, we denote $\mathbb{E}(W\mid\mathbf{x})$ as $w$, $\mathbb{E}(W\mid d, R=1)$ as $w^{+}_{d}$, $\mathbb{E}(W\mid d, R=0)$ as $w^{-}_{d}$ and $\Pr(R=1\mid\mathbf{x})$ as $p^{r}_{\mathbf{x}}$ in future formulation. Then we have:
\begin{equation}
\label{eq: watch_time_decompose}
    w = p^{r}_{\mathbf{x}}w^{+}_{d} + (1 - p^{r}_{\mathbf{x}})w^{-}_{d}.
\end{equation}
Eq.~\eqref{eq: watch_time_decompose} provides a unified formulation of duration bias and noisy watching rather than treating them as two separate mechanisms, which is beneficial for developing a unified method for addressing them simultaneously.
Based on decomposition on Eq.~\eqref{eq: watch_time_decompose}, we next give the error analysis of watch time as follows:

\begin{theorem}[Error of watch time] 
\label{thm: error of watch time}

for a given $(u,v)$, the error between scaled watch time $\frac{w}{w_{\mathrm{max}}}$ and its unobserved interest probability $p^{r}_{\mathbf{x}}$ is:
\begin{eqnarray*}
\left |\frac{w}{w_{\mathrm{max}}} - p^{r}_{\mathbf{x}}\right | = \left\lvert\frac{w^{+}_{d}-w_{\mathrm{max}}}{w_{\mathrm{max}}}p^{r}_{\mathbf{x}} + \frac{w^{-}_{d}}{w_{\mathrm{max}}}(1-p^{r}_{\mathbf{x}})\right \rvert\\ \leq \underbrace{\frac{w_{\mathrm{max}-w^{+}_{d}}}{w_{\mathrm{max}}}}_{\text{error of duration bias}}p^{r}_{\mathbf{x}} 
+ \underbrace{\frac{w^{-}_{d}}{w_{\mathrm{max}}}}_{\text{error of noisy watching}}(1-p^{r}_{\mathbf{x}}).
\end{eqnarray*}
\end{theorem}

The proof of the Theorem is apparent based on Eq.~\eqref{eq: watch_time_decompose}. As illustrated in Theorem~\ref{thm: error of watch time}, the upper bound of watch time's error can be divided as the linear combination of the error caused by duration bias and the error caused by noisy watching. The total error of watch time can be further reduced when both two errors are reduced. This error analysis proved the need to develop an approach to address both duration bias and noisy watching. 

% Meanwhile, we have the following corollary:

% \begin{corollary}[Effect of error] 
% \label{cor: trend of error}
% % $
% % % \mathbb{V}[\mathcal{L}_{\mathrm{AC}}] > \mathbb{V}[\mathcal{L}_{\mathrm{exam}}].
% % \mathbb{V}_\mathrm{AC} \geq \mathbb{V}_\mathrm{exam}.
% % $
% For watch time, the error of duration bias is mainly affect short video while the error of noisy watching is mainly affect long video.
% \begin{proof}
% since $w^{+}_{d}$ and $w^{-}_{d}$ is the monotonically increasing functions of $d$, then we can find that $\frac{w_{\mathrm{max}}-w^{+}_{d}}{w_{\mathrm{max}}}$ is the monotonically decreasing functions of $d$ while $\frac{w^{-}_{d}}{w_{\mathrm{max}}}$ is the monotonically increasing functions of $d$, thus proving the corollary.
% \end{proof}
% \end{corollary}
% In section~\ref{sec: result}, the experiment results will also verify this corollary.

\subsection{Analysis of Existing Methods}
\label{subsec: baseline analysis}
Methods have been proposed to address the issue brought by the duration bias, including Play Complete Rate, Watch Time Gain~\citep{Zheng2022DVR} and Duration-Deconfounded Quantile-based Method~\citep{Zhan2022Deconfounding}. However, the noisy watching is usually overlooked in these methods. Moreover, these methods uncover users' true interests with some critical assumptions on the user interest distribution, which are not always true in the real world, as shown in the following sections.
\subsubsection{Play Complete Rate}
In fact, the problem brought by duration bias is the different magnitude of different duration levels. In order to mitigate the effect of the magnitude, one direct idea is to scale each watch time $w$ with its corresponding video duration $d$ and employ this ratio as a surrogate label of user interest, which is called Play Complete Rate (PCR). For a given $(u,v)$, its PCR is formulated as:
\begin{equation}
\label{eq: pcr}
r^{\mathrm{PCR}}_{\mathbf{x}} = \frac{w}{d}.
\end{equation}
Compared with naively adopting watch time as the indicator of user interest, PCR takes a step towards scaling the magnitude according to each duration group and achieves better results. However, it can be shown that PCR can uncover user interest from watch time if and only if $w^{+}_{d}=C_{1}d$ and $w^{-}_{d}=C_{2}d$, where $C_1$ and $C_2$ are two constants. The detailed analysis can be found in Appendix~\ref{proof of prop pcr}

% PCR is far from uncovering user interest in the following proposition.

% \begin{proposition}[the assumption on which PCR relies] 
% \label{prop: error of pcr}
% For a given $(u,v)$ with duration $d$, only when  $w^{+}_{d}=C_{1}d$ and $w^{-}_{d}=C_{2}d$ ($C_1,C_2$ is two constants), PCR can uncover user interest from watch time. 
% \end{proposition}

% The proof of Proposition~\ref{prop: error of pcr} can be found in Appendix~\ref{proof of prop pcr}.

In practice, the underlying assumptions of PCR can hardly be satisfied. As shown in Fig~\ref{fig: evidence of bias and noise}, the curve of $w^{+}_{d}$ and $w^{-}_{d}$ with duration $d$ is not a linear function. As a consequence, the performance of PCR cannot be guaranteed. % and may arouse other negative effects.

\subsubsection{Watch Time Gain}
Watch time gain (WTG)~\citep{Zheng2022DVR} is a newly proposed state-of-the-art method for eliminating the duration bias. The core idea of WTG is to conduct standardization after video duration grouping, thus scaling the magnitude of watch time in each duration into the same interval. For a given $(u,v)$, its WTG is formulated as:
\begin{equation}
\label{eq: wtg}
r^{\mathrm{WTG}}_{\mathbf{x}} = \frac{w - \mu_{w}(d)}{\sigma_{w}(d)},
\end{equation}
where $\mu_{w}(d)$ is the average watch time and $\sigma_{w}(d)$ is the standard deviation of watch time for the videos with duration $d$.
Different from PCR, which only considers the watch time magnitude of the current sample, WTG is a method that aims to get a relative score among each duration group, thus further reducing the influence of duration bias. However, it can be shown that WTG can uncover user interest from watch time if and only if the distribution of user interest at each duration has the same expectation and standard deviation. The detailed analysis can be found in Appendix~\ref{proof of prop wtg}.

% However, we will demonstrate that WTG still relays on some critcal assumptions in the following proposition.

% \begin{proposition}[the assumption on which WTG relies] 
% \label{prop: error of wtg}
% For a given $(u,v)$ with duration $d$, only if the distribution of user interest at each duration has the same expectation and standard deviation, WTG can uncover user interest from watch time. 
% \end{proposition}

% The proof of Proposition~\ref{prop: error of wtg} can be found in Appendix~\ref{proof of prop wtg}.

In fact, it is unreasonable to assume that every duration group has a consistent user interest distribution. As illustrated in Fig~\ref{fig:causal_graph}, both user interest $R$ and video duration $D$ are determined by the feature of $(u,v)$. Therefore, the distribution of $R$ and the distribution of $D$ are still correlated, which violates the assumption of WTG.

\subsubsection{Duration-Deconfounded Quantile-based Method}
Duration-Deconfounded Quantile-based Method (D2Q)~\citep{Zhan2022Deconfounding} is another state-of-the-art method for alleviating duration bias. Unlike WTG, D2Q transforms the original watch time into the quantile score in each equal-frequency duration bin. For a given $(u,v)$, its D2Q label is formulated as:
\begin{equation}
\label{eq: D2Q}
r^{\mathrm{D2Q}}_{\mathbf{x}} = \frac{|\mathcal{D}| - M\pi_{m}(w)}{|\mathcal{D}|},
\end{equation}
where $|\mathcal{D}|$ is the total number of samples in the whole dataset, $M$ is the number of equal-frequency duration bins, $\pi_{m}(w): \mathbb{R}\rightarrow \{1, 2, \cdots, \frac{|\mathcal{D}|}{M}\}$ is a descending ranking function of watch time for current bin $m$. Similar to WTG, D2Q is also a kind of method for obtaining relative scores among each bin group. However, it can be shown that D2Q can uncover user interest from watch time if and only if all bins have the same ranking function of user interest. See the analysis in Appendix~\ref{proof of prop d2q} for the details.

% but it still requires some assumptions to guarantee the unbiasedness.

% \begin{proposition}[the assumption on which D2Q relies] 
% \label{prop: error of d2q}
% For a given $(u,v)$ with duration $d$, only if all bins have the same ranking function of user interest, D2D can uncover user interest from watch time. 
% \end{proposition}

% The proof of Proposition~\ref{prop: error of d2q} can be found in Appendix~\ref{proof of prop d2q}.

In order to hold the condition, it is necessary to reduce the number of bins, which in turn reduces the performance of debiasing. Moreover, the assumption is difficult to be tested in most cases.

\section{Our Approach: D$^2$Co}
To jointly mitigate the duration bias and noisy watching and relax the above assumptions, we propose Debiased and Denoised watch time Correction (D$^2$Co). Specifically, we first employ a duration-wise Gaussian Mixture Model plus frequency-weighted moving average for estimating the bias and noise terms. Then, we utilize a sensitivity-controlled correction function to separate user interest from watch time, which can reduce the sensitivity to estimation error.

\subsection{Estimating the Bias and Noise Terms}
% w = p^{r}_{\mathbf{x}}w^{+}_{d} + (1 - p^{r}_{\mathbf{x}})w^{-}_{d}
As illustrated in Eq.~\eqref{eq: watch_time_decompose}, the expected watch time $w$ of a given $(u,v)$ can be decomposed as the mixture of duration-biased watch time $w^{+}_{d}$ and noisy watch time $w^{-}_{d}$. From the perspective of probability, the distribution of watch time $\Pr(W=w\mid\mathbf{x})$ for a given $(u,v)$ can also be considered as the mixture of two latent distributions: $\Pr(W=w\mid d, R=1)$ and $\Pr(W=w\mid d, R=0)$, which is formulated as follows:
\begin{equation}
\label{eq: individual_gmm}
    % p^{w}_{\mathbf{x}}= p^{r}_{\mathbf{x}} p^{w^{+}}_{d}  + (1 - p^{r}_{\mathbf{x}})p^{w^{-}}_{d}
    \Pr(W=w\mid\mathbf{x}) = \sum_{R\in \{0,1\}}\Pr(R\mid\mathbf{x})\Pr(W=w\mid d,R),
\end{equation}
where $\Pr(W=w\mid d, R=1)$ is the distribution of the watch time due to the user's interest in videos with duration $d$, which suffers duration bias; $\Pr(W=w\mid d, R=0)$ is the distribution of the watch time that user watches videos with duration $d$ they are not interested in, which is controlled by noisy watching. The weight of each component is the user interest probability $\Pr(R=1|\mathbf{x})$ and $\Pr(R=0|\mathbf{x})$. 

\begin{figure*}
    \subfigure[Watch time dist. of different duration value]{
    \includegraphics[width=0.69\textwidth]{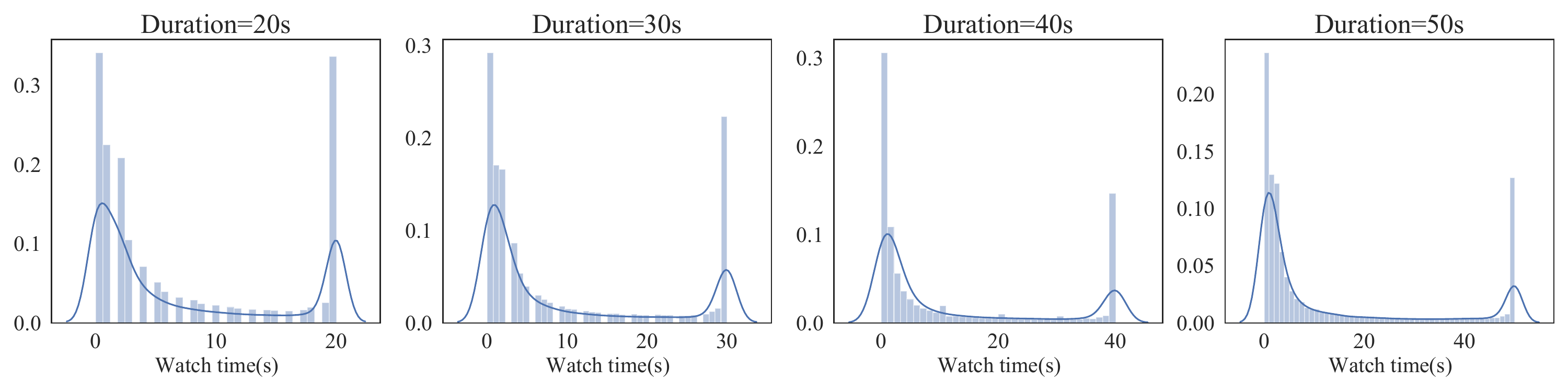}}
    \quad
    \subfigure[Watch time dist. of duration range]{
    \includegraphics[width=0.21\textwidth]{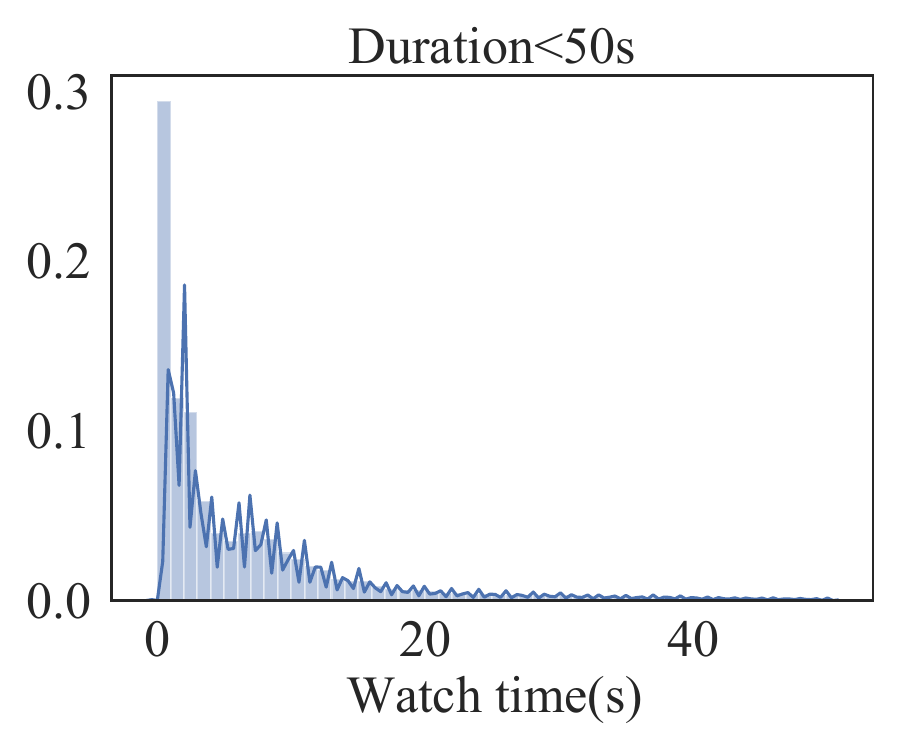}}
    \caption{The distribution of watch time in a different subset of KuaiRand.}
    \label{fig: dist of watch time}
\end{figure*}

To uncover user interest from the watch time, we need to estimate the parameters of the latent distributions. Here, we assume that $\Pr(W=w\mid d, R=1)$ and $\Pr(W=w\mid d, R=0)$ are two latent Gaussian distributions, which is a wild assumption. Then the Gaussian Mixture Model (GMM) can be utilized for estimating the parameters of latent mixture Gaussian distribution. However, Eq.~\eqref{eq: individual_gmm} lies on the individual level, which means we don't have enough samples to estimate the parameters of GMM for each individual. To this end, we transform the individual-level GMM equivalently to the duration level:
\begin{equation}
\label{eq: group_gmm}
\begin{split}
    % p^{w}_{\mathbf{x}}= p^{r}_{\mathbf{x}} p^{w^{+}}_{d}  + (1 - p^{r}_{\mathbf{x}})p^{w^{-}}_{d}
    \Pr(W=w\mid d) &= \sum_{\mathbf{x}}\Pr(\mathbf{x}\mid d)\Pr(W=w\mid d,\mathbf{x})\\
    &= \sum_{\mathbf{x}\in \mathcal{X}_d}\Pr(\mathbf{x})\Pr(W=w\mid\mathbf{x}) \\
    &= \sum_{\mathbf{x}\in \mathcal{X}_d}\Pr(\mathbf{x})\left( \sum_{R\in \{0,1\}}\Pr(R\mid\mathbf{x})\Pr(W=w\mid d,R)\right) \\ &
    = \sum_{R\in \{0,1\}} \left(\sum_{\mathbf{x}\in \mathcal{X}_d}\Pr(\mathbf{x})\Pr(R\mid\mathbf{x})\right) \Pr(W=w\mid d,R).
\end{split}
\end{equation}
Here, $\sum_{\mathbf{x}\in \mathcal{X}_d}\Pr(\mathbf{x})\Pr(R\mid\mathbf{x})$ can be regarded as the average user interest in videos of duration $d$. We can find that the latent distributions $\Pr(W=w\mid d, R=1)$ and $\Pr(W=w\mid d, R=0)$ are still the same as Eq.~\eqref{eq: individual_gmm}. As a result, we can estimate GMM parameters at the duration-level. To verify the rationality of the adoption of duration-level GMM, we show statistics on the distribution of watch time on the KuaiRand dataset. Fig.~\ref{fig: dist of watch time}(a) shows the watch time distribution of different duration groups(e.g.,  $Duration=20s,30s,40s,50s$). A significant bimodal phenomenon appears on those hist diagrams. However, as shown in Fig.~\ref{fig: dist of watch time}(b), this bimodal phenomenon disappears if we go to the watch time distribution of duration range (e.g.,  $Duration<50s$). This supports the rationality of regarding the watch time distribution as a mixture distribution in the duration-level.

Furthermore, considering that adjacent duration should have similar duration-biased watch time and noisy watch time, we employ a bi-directional frequency-weighted moving average to smooth the estimated sequence of duration-biased watch time $\hat{w}^{+}_{d}$ and noisy watch time $\hat{w}^{-}_{d}$. That is:
\begin{equation}
\label{eq: moving_ave}
\begin{split}
    % p^{w}_{\mathbf{x}}= p^{r}_{\mathbf{x}} p^{w^{+}}_{d}  + (1 - p^{r}_{\mathbf{x}})p^{w^{-}}_{d}
    \widetilde{w}^{+}_{d_i} &= \frac{|\mathcal{D}_{i-T}|\hat{w}^{+}_{d_{i-T}}+\cdots+|\mathcal{D}_i|\hat{w}^{+}_{d_i}+\cdots+|\mathcal{D}_{i+T}|\hat{w}^{+}_{d_{i+T}}}{|\mathcal{D}_{i-T}|+\cdots+|\mathcal{D}_i|+\cdots+|\mathcal{D}_{i+T}|}, \\
    \widetilde{w}^{-}_{d_i} &= \frac{|\mathcal{D}_{i-T}|\hat{w}^{-}_{d_{i-T}}+\cdots+|\mathcal{D}_i|\hat{w}^{-}_{d_i}+\cdots+|\mathcal{D}_{i+T}|\hat{w}^{-}_{d_{i+T}}}{|\mathcal{D}_{i-T}|+\cdots+|\mathcal{D}_i|+\cdots+|\mathcal{D}_{i+T}|},
\end{split}
\end{equation}
where $T$ denotes the window size of moving average. The smoothed $\widetilde{w}^{+}_{d}$ and $\widetilde{w}^{-}_{d}$ are leveraged to separate user interest from watch time in the next section.

\subsection{Separating User Interest from Watch Time}
\label{subsec: Separating_user_interest}
Based on Eq.~\eqref{eq: watch_time_decompose}, we can obtain the user interest with the bias term $\widetilde{w}^{+}_{d}$ and noise term $\widetilde{w}^{-}_{d}$ via affine correction, which named as D$^2$Co(A):
\begin{equation}
\label{eq: affine_D$^2$Co}
    % p^{w}_{\mathbf{x}}= p^{r}_{\mathbf{x}} p^{w^{+}}_{d}  + (1 - p^{r}_{\mathbf{x}})p^{w^{-}}_{d}
    r^{\mathrm{D^{2}Co(A)}}_{\mathbf{x}} (w,\widetilde{w}^{+}_{d},\widetilde{w}^{-}_{d}) = \frac{w - \widetilde{w}^{-}_{d}}{\widetilde{w}^{+}_{d} - \widetilde{w}^{-}_{d}}.
\end{equation}

If we estimate both the bias term and noise term accurately, Eq.~\eqref{eq: affine_D$^2$Co} undoubtedly equals user interest. As shown in the following theorem:

\begin{theorem}[Unbiasedness] 
\label{thm: unbiasedness of D$^2$Co(A)}
% $
% % \mathbb{V}[\mathcal{L}_{\mathrm{AC}}] > \mathbb{V}[\mathcal{L}_{\mathrm{exam}}].
% \mathbb{V}_\mathrm{AC} \geq \mathbb{V}_\mathrm{exam}.
% $
Given $(u,v)$, $r^{\mathrm{D^{2}Co(A)}}_{\mathbf{x}}$ is unbiased if the bias and noise terms are accurately estimated:
\[
r^{\mathrm{D^{2}Co(A)}}_{\mathbf{x}}(w,\widetilde{w}^{+}_{d},\widetilde{w}^{-}_{d}) \equiv p^{r}_{\mathbf{x}},\quad \mathrm{if}~ ~ \widetilde{w}^{+}_{d} = w^{+}_{d} \land \widetilde{w}^{-}_{d} = w^{-}_{d}.
\]
\end{theorem}
On the basis of Eq.~\eqref{eq: watch_time_decompose}, the proof of this theorem is apparent.

However, we can hardly accurately estimate the bias and noise term in practice. Once the estimation error occurs, then the above theorem will not hold. To this end, we analyzed the parameter sensitivity of $r^{\mathrm{D^{2}Co(A)}}_{\mathbf{x}}$ towards $\widetilde{w}^{+}_{d}$ and $\widetilde{w}^{-}_{d}$ respectively, which is given by the following theorem:

% \subsection{Discussion (Theoretical Analysis)}

\begin{theorem}[Parameter Sensitivity] 
\label{thm: parameter sensitivity of D$^2$Co(A)}
% $
% % \mathbb{V}[\mathcal{L}_{\mathrm{AC}}] > \mathbb{V}[\mathcal{L}_{\mathrm{exam}}].
% \mathbb{V}_\mathrm{AC} \geq \mathbb{V}_\mathrm{exam}.
% $
For a given disturbance (i.e., estimation error) $\delta_{\widetilde{w}^{+}_{d}}$ and $\delta_{\widetilde{w}^{-}_{d}}$ of the predict value of $\widetilde{w}^{+}_{d}$ and $\widetilde{w}^{-}_{d}$, if $w\in[\widetilde{w}^{-}_{d},\widetilde{w}^{+}_{d}]$, the sensitivity of $r^{\mathrm{D^{2}Co}}_{\mathbf{x}}$ to $\widetilde{w}^{+}_{d}$ and $\widetilde{w}^{-}_{d}$ is:
% \[
% \begin{split}
%     \mathbb{S}_{w^{+}_{d}} &> \mathbb{S}_{w^{-}_{d}},\quad \mathrm{if} ~ ~ w\in(\frac{w^{+}_{d}+w^{-}_{d}}{2}, w^{+}_{d}]\\
%     \mathbb{S}_{w^{+}_{d}} &\leq \mathbb{S}_{w^{-}_{d}},\quad \mathrm{if} ~ ~ w\in[w^{-}_{d},\frac{w^{+}_{d}+w^{-}_{d}}{2}]
% \end{split}
% \]
\[
\begin{split}
    \mathbb{S}_{\widetilde{w}^{+}_{d}} &= \left | \frac{\partial~r^{\mathrm{D^{2}Co}}_{\mathbf{x}}(w,\widetilde{w}^{+}_{d},\widetilde{w}^{-}_{d})}{\partial~ \widetilde{w}^{+}_{d}} \delta_{\widetilde{w}^{+}_{d}}\right| = \frac{w-\widetilde{w}^{-}_{d}}{(\widetilde{w}^{+}_{d} - \widetilde{w}^{-}_{d})^2} \left|\delta_{\widetilde{w}^{+}_{d}}\right|, \\
    \mathbb{S}_{\widetilde{w}^{-}_{d}} &= \left | \frac{\partial~r^{\mathrm{D^{2}Co}}_{\mathbf{x}}(w,\widetilde{w}^{+}_{d},\widetilde{w}^{-}_{d})}{\partial~ \widetilde{w}^{-}_{d}} \delta_{\widetilde{w}^{-}_{d}}\right| = \frac{\widetilde{w}^{+}_{d}-w}{(\widetilde{w}^{+}_{d} - \widetilde{w}^{-}_{d})^2} \left|\delta_{\widetilde{w}^{-}_{d}}\right| ,
    % \mathbb{S}_{w^{+}_{d}} &= \left|\frac{w-w^{-}_{d}}{w^{+}_{d}-w^{-}_{d}} - \frac{w-w^{-}_{d}}{w^{+}_{d} + \delta - w^{-}_{d}}\right| = \frac{w-w^{-}_{d}}{w^{+}_{d}-w^{-}_{d}}\left|\frac{\delta}{w^{+}_{d}-w^{-}_{d}+\delta}\right| \\
    % % \left|\frac{w-w^{-}_{d}}{w^{+}_{d}-w^{-}_{d}} - \frac{w-w^{-}_{d}}{w^{+}_{d} + \delta - w^{-}_{d}}\right| = \left |\frac{\delta}{w^{+}_{d}-w^{-}_{d}}\left(\frac{w-w^{-}_{d}}{w^{+}_{d}-w^{-}_{d}+\delta}\right) \right | \\
    % \mathbb{S}_{w^{-}_{d}} &= \left|\frac{w-w^{-}_{d}}{w^{+}_{d}-w^{-}_{d}} - \frac{w - w^{-}_{d} - \delta}{w^{+}_{d} - w^{-}_{d} - \delta}\right| = \frac{w^{+}_{d}-w}{w^{+}_{d}-w^{-}_{d}}\left | \frac{\delta}{w^{+}_{d}-w^{-}_{d}-\delta}\right |
    % % \mathbb{S}_{w^{-}_{d}} &= \left|\frac{w-w^{-}_{d}}{w^{+}_{d}-w^{-}_{d}} - \frac{w - w^{-}_{d} - \delta}{w^{+}_{d} - w^{-}_{d} - \delta}\right| = \left | \frac{\delta}{w^{+}_{d}-w^{-}_{d}}\left(\frac{w^{+}_{d}-w}{w^{+}_{d}-w^{-}_{d}-\delta}\right) \right |
\end{split}
\]
where $\mathbb{S}_{\widetilde{w}^{+}_{d}}$ and $\mathbb{S}_{\widetilde{w}^{-}_{d}}$ is the sensitivity of $r^{\mathrm{D^{2}Co}}_{\mathbf{x}}$ to $\widetilde{w}^{+}_{d}$ and $\widetilde{w}^{-}_{d}$ respectively.
\end{theorem}
% The proof Theorem~\ref{thm: parameter sensitivity of D$^2$Co(A)} can be found in Appendix~\ref{proof of sensitivity}.
The proof of Theorem~\ref{thm: parameter sensitivity of D$^2$Co(A)} is based on the definition of parameter sensitivity. This theorem indicates that the estimation error of bias and noise terms has different effects at different watch time. For $\mathbb{S}_{w^{+}_{d}}$, it has large value with the growth of $w$. In contrast, $\mathbb{S}_{w^{-}_{d}}$ has lower value with the growth of $w$. From the perspective of the entire dataset, the dataset with the majority of short watch time is mainly affected by $w^{-}_{d}$. In contrast, the dataset with the majority of long watch time is mainly affected by $w^{+}_{d}$. To this end, we proposed a sensitivity-controlled correction function that adjusts sensitivity preferences according to the proportion of watch time in the dataset:
\begin{equation}
\label{eq: sense_control_D$^2$Co}
    % p^{w}_{\mathbf{x}}= p^{r}_{\mathbf{x}} p^{w^{+}}_{d}  + (1 - p^{r}_{\mathbf{x}})p^{w^{-}}_{d}
    r^{\mathrm{D^{2}Co(S)}}_{\mathbf{x}}(w,\widetilde{w}^{+}_{d},\widetilde{w}^{-}_{d})= \frac{\exp(\alpha w) - \exp(\alpha\widetilde{w}^{-}_{d})}{\exp(\alpha\widetilde{w}^{+}_{d}) - \exp(\alpha\widetilde{w}^{-}_{d})},
\end{equation}
where $\alpha$ is the sensitivity control term. We can prove that, $r^{\mathrm{D^{2}Co(S)}}_{\mathbf{x}}$ has a lower sensitivity to parameters $w^{+}_{d}$ and $w^{-}_{d}$ compared to $r^{\mathrm{D^{2}Co(A)}}_{\mathbf{x}}$ through the following proposition.

\begin{proposition}[D$^2$Co(S) has lower sensitivity] 
\label{prop: lower sensitivity of D$^2$Co(S)}
% $
% % \mathbb{V}[\mathcal{L}_{\mathrm{AC}}] > \mathbb{V}[\mathcal{L}_{\mathrm{exam}}].
% \mathbb{V}_\mathrm{AC} \geq \mathbb{V}_\mathrm{exam}.
% $
For a given $(u,v)$, denoting the sensitivity of D$^2$Co(S) as $\mathbb{S}^{\prime}_{w^{+}_{d}}$ and $\mathbb{S}^{\prime}_{w^{-}_{d}}$, we have:
\[
\begin{split}
    \mathbb{S}^{\prime}_{w^{+}_{d}}&<\mathbb{S}_{w^{+}_{d}}, \quad\mathrm{if} ~ ~ \alpha<0 ,\\
    \mathbb{S}^{\prime}_{w^{-}_{d}}&<\mathbb{S}_{w^{-}_{d}}, \quad\mathrm{if} ~ ~ \alpha>0 .
\end{split}
\]
\end{proposition}
Due to the limitation of the page, proof Proposition~\ref{prop: lower sensitivity of D$^2$Co(S)} can be found in supplementary material. %anonymous website\footnote{\url{https://github.com/hyz20/D2Co.git}}. 
In practice, we need to tune the value of $\alpha$ for controlling the sensitivity of D$^2$Co(S) towards $w^{+}_{d}$ and $w^{-}_{d}$. 

The pipeline of our method is shown in Algorithm~\ref{alg: D$^2$Co}. In summary, we employ a duration-wise Gaussian Mixture Model and a frequency-weighted moving average to estimate the bias and noise terms. Then, we utilize a sensitivity-controlled correction function instead of a standard affine correction to better separate user interest from watch time. The separated user interest can be utilized as the supervision signal for learning a better recommendation model.

\begin{algorithm}[t]
    \SetAlgoLined
    \KwIn{User interactions $\mathcal{D}=\{(\mathbf{x}_i, w_{i}, d_{i})\}_{i=1}^{N}$, moving average windows size $T$, sensitivity control term $\alpha$}
    $\mathbf{W}^+ \leftarrow \{\;\}$, $\mathbf{W}^- \leftarrow \{\;\}$ , $\mathbf{\mathcal{R}} \leftarrow \{\;\}$\;
    % \tcp{For each duration second}
    % \For{\xujun{$d \in \mathcal{D}$}  }
    % \For{$d\in [d_{\mathrm{min}}, d_{\mathrm{max}}]$}
    \For{$d \in \{d_{\mathrm{min}},\cdots, d_{\mathrm{max}}\}$}
    {
%        $\mathcal{D}^{\prime} = \{(\mathbf{x}_i, w_{i}, d_{i})_{i=1}^{N^{\prime}}\mid\mathbb{I}(d_{i}=d)\}$ \;
         % \xujun{$\mathcal{D}^{\prime} = \{(\mathbf{x}_i, w_{i}, d_{i})\mid (\mathbf{x}_i, w_{i}, d_{i})\in \mathcal{D} \wedge (d_{i}=d)\}$ \;} % 从D中选出d' = d的那些数据，是这个逻辑吗？ 是的
        $\mathcal{D}^{\prime} = \{(\mathbf{x}_i, w_{i}, d_{i})\mid (\mathbf{x}_i, w_{i}, d_{i})\in \mathcal{D} \wedge (d_{i}=d)\}$ \;
        $\mathbf{W}^+[d], \mathbf{W}^-[d]\leftarrow \mathrm{GMM}(\mathcal{D}^{\prime}, \mathrm{components}=2)$ \;
    
    }
    $\widetilde{\mathbf{W}}^+ \leftarrow \mathrm{Moving\_Average}(\mathbf{W}^+, T)$ ~(Eq.\eqref{eq: moving_ave})\;
    $\widetilde{\mathbf{W}}^- \leftarrow \mathrm{Moving\_Average}(\mathbf{W}^-, T)$~(Eq.\eqref{eq: moving_ave}) \;

    % \tcp{For each data sample}
    \For{$(\mathbf{x}_i, w_{i}, d_{i}) \in \mathcal{D}$} 
    {
        $\mathbf{\mathcal{R}}[i] \leftarrow r^{\mathrm{D^{2}Co(S)}}_{\mathbf{x}}(w_i, \widetilde{\mathbf{W}}^{+}[d_i],\widetilde{\mathbf{W}}^{-}[d_i], \alpha)$  ~(Eq.~\eqref{eq: sense_control_D$^2$Co}) \;
    }
    \Return{ $\mathbf{\mathcal{R}}$}
    \caption{The pipeline of D$^2$Co}
    \label{alg: D$^2$Co}
\end{algorithm}

\section{Experiments and Results}

\begin{table}[t]
\caption{Statistics of the datasets adopted in this study}
\resizebox{0.45\textwidth}{!}{
\begin{tabular}{ccccc}
\hline
Dataset  & \#Users & \#Videos & \#Interactions & Duration Ranges(s) \\ \hline
KuaiRand & 26,988  & 6,598    & 1,266,560      & {[}5,240{]}     \\
WeChat   & 20,000  & 96,418   & 7,310,108      & {[}5,60{]}      \\ \hline
\end{tabular}}
\label{tab: dataset statistics}
\end{table}

\subsection{Experimental setting}
\subsubsection{Datasets}
For evaluating the performance of proposed D$^2$Co, we utilize two public real-world datasets: WeChat\footnote{\url{https://algo.weixin.qq.com/}} and KuaiRand\footnote{\url{http://kuairand.com/}}. They are respectively collected from two large micro-video platforms, Wechat Channels and Kuaishou. We list their statistic information in Table~\ref{tab: dataset statistics}. The details of these two datasets are as follows:

\textbf{WeChat}. This dataset is released by WeChat Big Data Challenge 2021, containing the Wechat Channels logs within two weeks. Following the practice in \citep{Zheng2022DVR}, we split the data into the first 10 days, the middle 2 days, and the last 2 days as training, validation, and test set. The adopted input features include \emph{userid},\emph{feedid},\emph{device},\emph{authorid},\quad\emph{bgm\_song\_id},\emph{bgm\_singer\_id},\emph{user\_type}, \emph{like}, \emph{read\_comment}, \emph{forward}.

\textbf{KuaiRand}~\citep{gao2022kuairand}. KuaiRand is a newly released sequential recommendation dataset collected from KuaiShou. As suggested in~\citep{gao2022kuairand}, we utilized one of the subsets \emph{KuaiRand-pure} in this study. To mitigate the sparsity problem, we selected data from which the video duration is up to 4 minutes. We split the data into the first 14 days, the middle 7 days, and the last 10 days as training, validation, and test set. The adopted input features include \emph{user\_id}, \emph{video\_id}, \emph{author\_id}, \emph{music\_id}, \emph{video\_type},\emph{upload\_type},\emph{tab},\emph{is\_like}, \emph{is\_follow}, \emph{is\_comment}, \emph{is\_forward}, \emph{is\_profile\_enter},\emph{is\_hate}, \emph{most\_popular\_tag}.

% \begin{itemize}
%     \item 
%     \item 
% \end{itemize}
\subsubsection{Evaluation}
As we discussed before, the watch time is an unreliable label for measuring user interest. For evaluating the performance of mitigating the duration bias and noisy watching in watch time, we need to know the true user interest in the recommended video. Since the explicit feedbacks suffer the spareness problem, we cannot directly utilize them as ground truth labels in our experiments. 
% Inspired by the definition of \emph{long\_view} in KuaiRand dataset. 
To this end, we adopt the definition of \emph{long\_view} from the KuaiRand dataset~\citep{gao2022kuairand} as the user interest indicator, which defines the user interest for a given $(u,v)$ as follows:

\begin{equation}
    \begin{split}
        r_{\mathbf{x}} = \left\{
        \begin{aligned}
            & 1,\quad \mathrm{if}~~ (d\leq18s \land w=d) \lor (d>18s \land w>18s) \mathrm{;} \\
            % & 1,\quad \mathrm{if}~~ d>18s \land w>18s \mathrm{;}  \\
            & 0,\quad \mathrm{else;} 
        \end{aligned}
        \right.
    \end{split}
    \label{eq: interest define}
\end{equation}

It is worth noting that this kind of definition is close to Valid Viewing (VV), which is one of the online metrics we leveraged in online A/B testing (section~\ref{sec: ab_test}).
Unlike the RMSE used in \citep{Zhan2022Deconfounding} and WTG used in \citep{Zheng2022DVR}, we are mainly concerned about whether the recommendation model can rank interesting videos in top-ranking positions, so the GAUC and nDCG@k are utilized as the evaluation metric of recommendation performance.

\subsubsection{Baselines}
% For addressing the issue brought by duration bias, some baselines have been proposed before.~\citep{Zhan2022Deconfounding,Zheng2022DVR} In section~\ref{subsec: baseline analysis}, we analyze these baselines and point out that that they each need to rely on critical assumptions to obtain user interest from the watch time. 
As have been described in Section~\ref{subsec: Separating_user_interest}, D$^2$Co has two versions \textbf{D$^2$Co(A)} and \textbf{D$^2$Co(S)}. In our experiments, we will compare our proposed method with these baselines: \textbf{PCR}, \textbf{D2Q}~\citep{Zhan2022Deconfounding} and \textbf{WTG}~\citep{Zheng2022DVR}. To investigate the generalization of our method and baselines, we integrate them with different backbone models. Specifically, we use the classical linear recommendation model \textbf{FM}~\citep{Rendle2012Factorization}, the classical deep recommendation model \textbf{DeepFM}~\citep{guo2017deepfm} and the state-of-the-art recommendation model \textbf{AutoInt}~\citep{Song2019AutoInt} as our backbone recommendation model.

Moreover, considering that the existing baselines overlook the noisy watching, we equip those baselines with denoise capability via data post-processing. Specifically, we treat all samples with less than 5 seconds of watch time as 0 values after calculating the value of baseline labels. This simple post-processing divides the noise samples by threshold so that the baselines have denoise capability, and they are denoted as \textbf{PCR-denoise}, \textbf{D2Q-denoise}, and \textbf{WTG-denoise}.

\subsubsection{Implementation Details}
We implement all the backbones with pytorch-fm\footnote{\url{https://github.com/rixwew/pytorch-fm}}, an open-source library for factorization machine models. We employ Binary Cross Entropy Loss for all baselines and our methods for fair comparisons. In particular, we transform WTG into probability via the cumulative density function $\Phi(\cdot)$ of standard Gaussian distribution. For D$^2$Co(A) and D$^2$Co(S), we clip their value into the interval $[0,1]$. For D2Q, the group number is set to 60 in KuaiRand and 30 in WeChat. We utilize Adam as the optimizer and set the initial learning rate as 0.001. The batch size is set as 512. For all the backbone models, we set their latent embedding dimension to 10. For all methods with neural networks, the hidden units are set to 64 while the dropout ratio is set to 0.2. The value of moving average window size $T$ is tuned in the interval $[1,5]$, and the value of sensitivity control term $\alpha$ is tuned in the interval $[1e^{-2},5e^{-2}]$ in WeChat dataset and $[-1e^{-2},-5e^{-2}]$ in KuaiRand dataset. We tune our hyper parameters on the validation set while evaluating the performance on the test set. The source code is available at \url{https://github.com/hyz20/D2Co.git}.

\subsection{Overall Performance}
% \usepackage{multirow}
% \usepackage{multirow}
% \usepackage{multirow}
% \usepackage{multirow}

\begin{table*}[t]
\caption{The recommendation performance of D$^2$Co and other baselines in KuaiRand and WeChat. \textbf{Boldface} means the best performed methods (excluding Oracle), while {\underline{underline}} means the second best performed methods, superscripts $\dag$ means the significance compared to the second best performed methods with $p<0.05$ of one-tailed $t$-test .}
\label{tab: main_result}
\resizebox{0.8\textwidth}{!}{
\begin{tabular}{cl|cccc|cccc}
\hline
\multicolumn{2}{c|}{}                                    & \multicolumn{4}{c|}{KuaiRand}                                     & \multicolumn{4}{c}{WeChat}                                                                                      \\ \hline
\multicolumn{1}{l|}{Backbone}              & Methods     & GAUC           & nDCG@1         & nDCG@3         & nDCG@5         & \multicolumn{1}{l}{GAUC} & \multicolumn{1}{l}{nDCG@1} & \multicolumn{1}{l}{nDCG@3} & \multicolumn{1}{l}{nDCG@5} \\ \hline
\multicolumn{1}{c|}{\multirow{10}{*}{FM}}  & Watch Time  & 0.584          & 0.402          & 0.461          & 0.501          & 0.506                    & 0.538                      & 0.542                      & 0.547                      \\
\multicolumn{1}{c|}{}                      & PCR         & 0.626          & 0.432          & 0.482          & 0.517          & 0.532                    & 0.557                      & 0.560                      & 0.565                      \\
\multicolumn{1}{c|}{}                      & PCR-denoise & 0.636          & 0.437          & 0.487          & 0.521          & 0.532                    & 0.560                      & 0.563                      & 0.567                      \\
\multicolumn{1}{c|}{}                      & D2Q         & 0.628          & 0.433          & 0.484          & 0.519          & 0.533                    & 0.546                      & 0.553                      & 0.560                      \\
\multicolumn{1}{c|}{}                      & D2Q-denoise & 0.641          & 0.441          & 0.490          & 0.524          & 0.538                    & 0.559                      & 0.563                      & 0.569                      \\
\multicolumn{1}{c|}{}                      & WTG         & 0.635          & 0.437          & 0.486          & 0.520          & 0.541                    & 0.556                      & 0.562                      & 0.569                      \\
\multicolumn{1}{c|}{}                      & WTG-denoise & {\underline{0.645}}          & {\underline{0.442}}          & {\underline{0.491}}          & {\underline{0.525}}          & {\underline{0.545}}                    & {\underline{0.564}}                      & {\underline{0.567}}                      & {\underline{0.572}}                      \\ \cline{2-10} 
\multicolumn{1}{c|}{}                      & D$^2$Co(A)     & 0.650          & 0.446          & 0.493          & 0.527          & 0.551                    & 0.577                      & 0.578                      & 0.583                      \\
\multicolumn{1}{c|}{}                      & D$^2$Co(S)     & $\textbf{0.653}^\dag$ & $\textbf{0.451}^\dag$ & $\textbf{0.497}^\dag$ & $\textbf{0.530}$ & $\textbf{0.556}^\dag$           & $\textbf{0.581}^\dag$             & $\textbf{0.586}^\dag$             & $\textbf{0.590}^\dag$             \\ \cline{2-10} 
\multicolumn{1}{c|}{}                      & Oracle      & 0.664          & 0.456          & 0.502          & 0.535          & 0.556                    & 0.585                      & 0.587                      & 0.590                      \\ \hline
\multicolumn{1}{c|}{\multirow{10}{*}{DeepFM}} & Watch Time  & 0.593          & 0.402          & 0.464          & 0.503          & 0.506                    & 0.554                      & 0.555                      & 0.560                      \\
\multicolumn{1}{c|}{}                      & PCR         & 0.628          & 0.435          & 0.483          & 0.518          & 0.531                    & 0.559                      & 0.562                      & 0.568                      \\
\multicolumn{1}{c|}{}                      & PCR-denoise & 0.637          & 0.440          & 0.488          & 0.523          & 0.532                    & 0.559                      & 0.562                      & 0.569                      \\
\multicolumn{1}{c|}{}                      & D2Q         & 0.635          & 0.437          & 0.489          & 0.522          & 0.532                    & 0.550                      & 0.554                      & 0.562                      \\
\multicolumn{1}{c|}{}                      & D2Q-denoise & 0.642          & 0.443          & 0.492          & 0.525          & 0.537                    & 0.564                      & 0.565                      & 0.572                      \\
\multicolumn{1}{c|}{}                      & WTG         & 0.635          & 0.436          & 0.486          & 0.520          & 0.542                    & 0.561                      & 0.564                      & 0.571                      \\
\multicolumn{1}{c|}{}                      & WTG-denoise & {\underline{0.647}}          & {\underline{0.444}}          & {\underline{0.493}}          & {\underline{0.526}}          & {\underline{0.544}}                    & {\underline{0.571}}                      & {\underline{0.570}}                      & {\underline{0.577}}                      \\ \cline{2-10} 
\multicolumn{1}{c|}{}                      & D$^2$Co(A)     & 0.653         & 0.447          & 0.496          & 0.528          & 0.551                    & 0.574                      & 0.576                      & 0.583             \\
\multicolumn{1}{c|}{}                      & D$^2$Co(S)     & $\textbf{0.656}^\dag$ & $\textbf{0.451}^\dag$ & $\textbf{0.499}^\dag$ & $\textbf{0.532}$ & $\textbf{0.555}^\dag$           & $\textbf{0.587}^\dag$             & $\textbf{0.587}^\dag$             & $\textbf{0.593}^\dag$             \\ \cline{2-10} 
\multicolumn{1}{c|}{}                      & Oracle      & 0.666          & 0.459          & 0.505          & 0.537          & 0.556                    & 0.583                      & 0.585                      & 0.591                      \\ \hline
\multicolumn{1}{c|}{\multirow{10}{*}{AutoInt}} & Watch Time  & 0.592          & 0.398          & 0.461          & 0.501          & 0.506                    & 0.559                      & 0.557                      & 0.562                      \\
\multicolumn{1}{c|}{}                      & PCR         & 0.624          & 0.429          & 0.480          & 0.515          & 0.532                    & 0.555                      & 0.559                      & 0.567                      \\
\multicolumn{1}{c|}{}                      & PCR-denoise & 0.639          & 0.441          & 0.489          & 0.524          & 0.533                    & 0.561                      & 0.563                      & 0.570                      \\
\multicolumn{1}{c|}{}                      & D2Q         & 0.633          & 0.436          & 0.486          & 0.521          & 0.535                    & 0.553                      & 0.556                      & 0.564                      \\
\multicolumn{1}{c|}{}                      & D2Q-denoise & 0.641          & 0.438          & 0.490          & 0.524          & 0.539                    & 0.563                      & 0.566                      & 0.573                      \\
\multicolumn{1}{c|}{}                      & WTG         & 0.637          & 0.437          & 0.487          & 0.521          & 0.544                    & 0.562                      & 0.563                      & 0.570                      \\
\multicolumn{1}{c|}{}                      & WTG-denoise & {\underline{0.645}}          & {\underline{0.441}}          & {\underline{0.491}}          & {\underline{0.525}}          & {\underline{0.547}}                    & {\underline{0.569}}                      & {\underline{0.571}}                      & {\underline{0.578}}                      \\ \cline{2-10} 
\multicolumn{1}{c|}{}                      & D$^2$Co(A)     & 0.653          & 0.448          & 0.496          & 0.529          & 0.551                    & 0.575                      & 0.578                      & 0.585            \\
\multicolumn{1}{c|}{}                      & D$^2$Co(S)     & $\textbf{0.658}^\dag$ & $\textbf{0.453}^\dag$ & $\textbf{0.499}^\dag$ & $\textbf{0.532}^\dag$ & $\textbf{0.556}^\dag$     & $\textbf{0.581}\dag$             & $\textbf{0.586}^\dag$             & $\textbf{0.593}^\dag$             \\ \cline{2-10} 
\multicolumn{1}{c|}{}                      & Oracle      & 0.665          & 0.459          & 0.502          & 0.536          & 0.557                    & 0.585                      & 0.587                      & 0.594                      \\ \hline
\end{tabular}}
\end{table*}

Table~\ref{tab: main_result} illustrates the recommendation performance of proposed D$^2$Co and other baselines. According to the result in Table~\ref{tab: main_result}, our proposed D$^2$Co(S) obtains the best performance on both KuaiRand and WeChat datasets and all backbones significantly. In addition, the recommendation models trained with debiased labels PCR, D2Q, and WTG outperform those trained by Watch Time by a large margin since they mitigate the duration bias. Then, Our proposed D$^2$Co(A) and D$^2$Co(S) further outperform these debias baselines since our proposed methods consider the noisy watching. Furthermore, our proposed D$^2$Co(S) has better performance than D$^2$Co(A) in both datasets. This shows the superiority of our sensitivity-controlled correction. In section~\ref{subsec: effect of sense control}, we will reveal the intrinsic reasons why D$^2$Co(S) exceeds D$^2$Co(A).

It is worth noting that those baselines equipped with denoise post-processing (PCR-denoise, D2Q-denoise, WTG-denoise) have different degrees of improvement compared to their original methods. 
% Especially for WTG, when equiped with denoise post-processing, it can reach the competitive performance with our D$^2$Co(A) but still perform worse than D$^2$Co(S). 
This phenomenon clearly confirms the existence of noisy watching. However, the denoise post-processing is just a heuristic truncation of the short watch time samples, which only removes part of the noise. 
% In Eq.~\eqref{eq: watch_time_decompose}, we assume that all the watch time length may contain noisy watching. 
Hence, there still exist performance gaps between D$^2$Co(S) and most denoised baselines. Moreover, the gap between original baselines and Watch Time is larger than that between D$^2$Co(S) and original baselines in KuaiRand dataset. Therefore, we can conclude that duration bias is more harmful than noisy watching in the KuaiRand dataset. In contrast, the gap between original baselines and Watch Time is smaller than that between D$^2$Co(S) and original baselines in WeChat dataset, which indicate that noisy watching is the main problem in this dataset. We will further discuss this conclusion in section~\ref{sec: bias and noise}.

\begin{figure*}
    \subfigure[KuaiRand]{
    \includegraphics[width=0.25\textwidth]{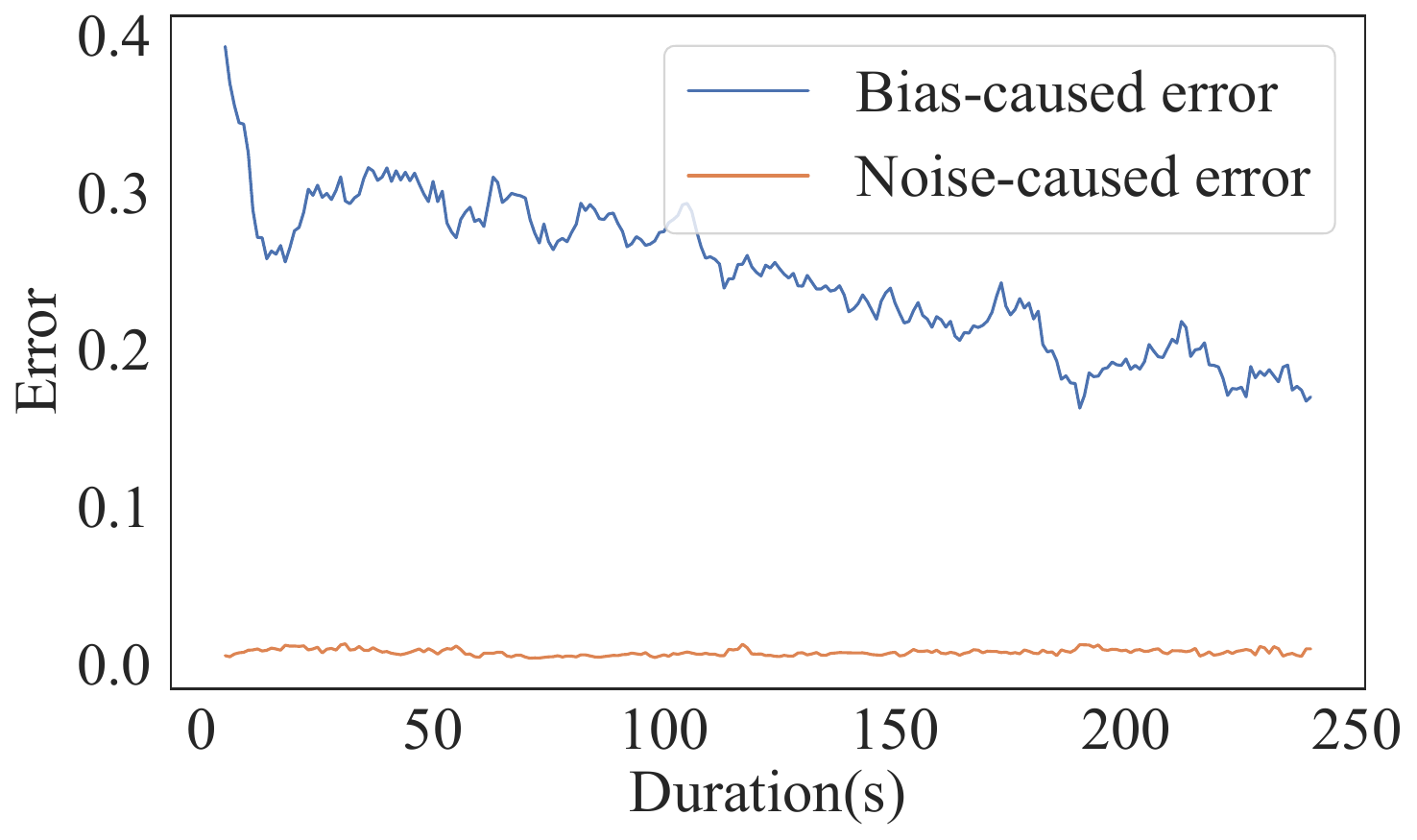}}
    \quad
    \subfigure[WeChat]{
    \includegraphics[width=0.25\textwidth]{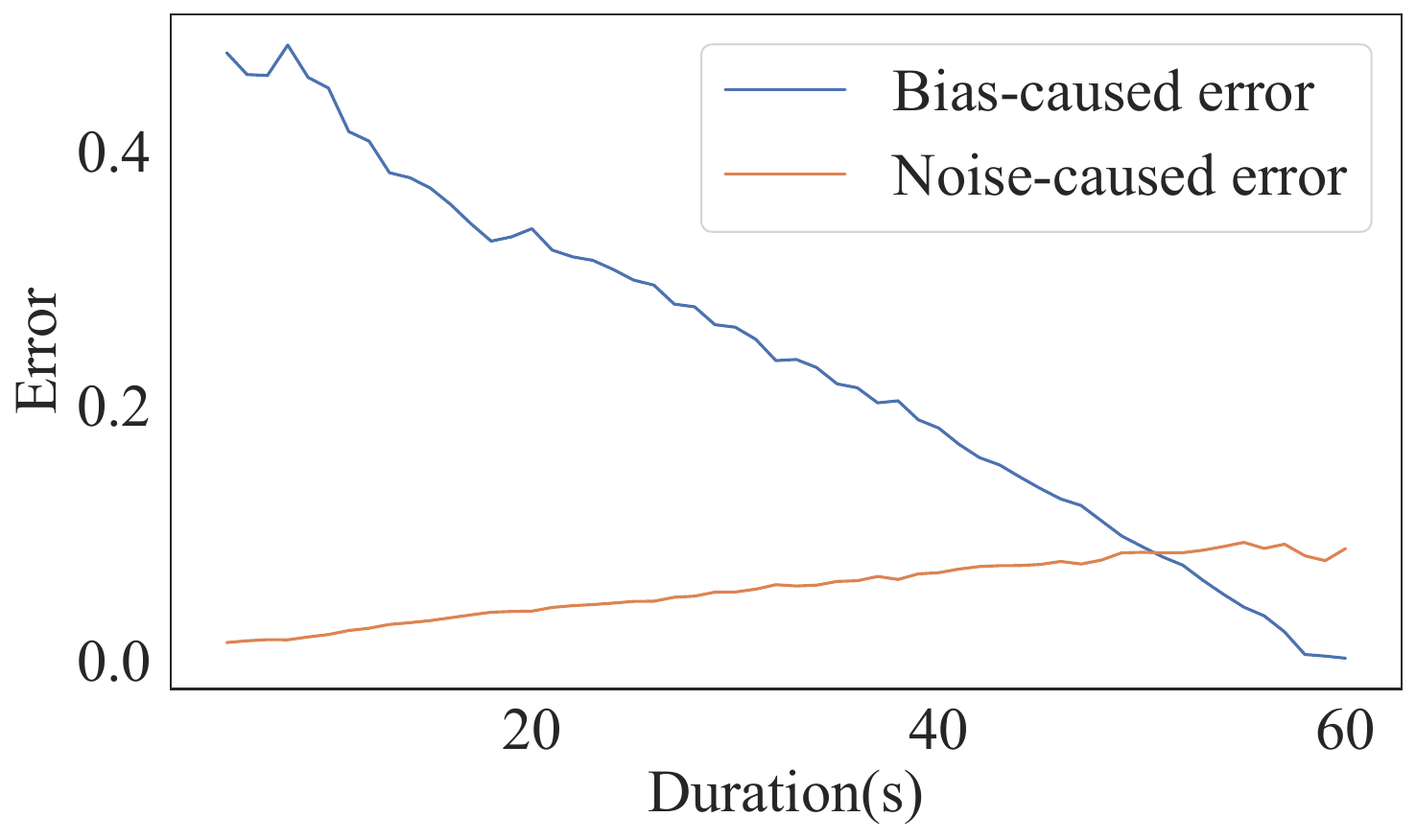}}
    \caption{The curve of the mean error caused by duration bias and noisy watching with the growth of duration, w.r.t KuaiRand and WeChat datasets.}
    \label{fig: bias_noise_curve}
\end{figure*}

\begin{table*}[t]
\caption{The nDCG@1 of D$^2$Co and other baselines in three equal frequency duration range. \textbf{Boldface} means the best performed methods (excluding Oracle), while {\underline{underline}} means the second best performed methods, superscripts $\dag$ means the significance compared to the second best performed methods with $p<0.05$ of one-tailed $t$-test . The backbone recommendation model is DeepFM. }
\resizebox{0.9\textwidth}{!}{
\begin{tabular}{c|c|c|ccc|cc|c}
\hline
Dataset                   & Duration Range & \multicolumn{1}{l|}{Watch Time} & PCR            & D2Q            & WTG            & \multicolumn{1}{l}{D$^2$Co(A)} & D$^2$Co(S)         & Oracle \\ \hline
\multirow{3}{*}{KuaiRand} & (0,32{]}       & 0.380                           & 0.389(+32.7\%) & 0.391(+36.4\%) & {\underline{0.391(+36.8\%)}} & $\textbf{0.397(+58.5\%)}^\dag$     & 0.397(+58.3\%)  & 0.409  \\
                          & (32,94{]}      & 0.394                           & 0.398(+20.6\%) & {\underline{0.406(+67.6\%)}} & 0.402(+46.9\%) & 0.409(+86.9\%)              & \textbf{0.411(+99.3\%)}  & 0.411  \\
                          & (94,240{]}     & 0.371                           & 0.374(+19.1\%) & 0.373(+10.0\%) & {\underline{0.375(+20.0\%)}} & 0.382(+58.6\%)              & $\textbf{0.389(+92.9\%)}^\dag$  & 0.390  \\ \hline
\multirow{3}{*}{WeChat}   & (0,16{]}       & 0.554                           & {\underline{0.573(+57.0\%)}} & 0.565(+31.6\%) & 0.569(+44.2\%) & 0.579(+74.3\%)              & $\textbf{0.591(+108.9\%)}^\dag$ & 0.588  \\
                          & (16,42{]}      & 0.549                           & {\underline{0.555(+28.4\%)}} & 0.545(-16.1\%) & 0.554(+22.9\%) & 0.568(+85.6\%)              & $\textbf{0.569(+91.5\%)}^\dag$  & 0.571  \\
                          & (42,60{]}      & 0.548                           & 0.546(-20.3\%) & 0.544(-35.4\%) & {\underline{0.548(-4.8\%)}}  & 0.556(+61.5\%)              & $\textbf{0.558(+70.7\%)}^\dag$  & 0.561  \\ \hline
\end{tabular}}
\label{tab: group_result}
\end{table*}

\subsection{The Effectiveness of Mitigating Bias and Noise}
\label{sec: bias and noise}
Although Tabel~\ref{tab: main_result} shows a significant improvement of our D$^2$Co compared to the baselines, it is still unclear how much of these improvements come from the denoise that we claim to have taken into account. In Theorem~\ref{thm: error of watch time}, we analyzed the error of watch time and divided the overall error into the duration bias-caused error and noisy watching-caused error. On this basis, we first present the curve of mean error with video duration in Fig.~\ref{fig: bias_noise_curve}, with the estimated $w^{+}_{d}$ and $w^{-}_{d}$. In Fig.~\ref{fig: bias_noise_curve}(a), the error caused by duration bias is much larger than that of noisy watching, and the curve of noisy watching is close to zero. This indicates that duration bias dominates the error of watch time in KuiaRand. In Fig.~\ref{fig: bias_noise_curve}(b), the error caused by noisy watching is an increasing curve, while the error caused by duration bias is a decreasing curve. This indicates that the duration bias dominates the overall error of watch time in short-duration intervals of WeChat. However, in long-duration intervals of WeChat, the noisy watching dominates the watch time's overall error.

% In Fig.~\ref{fig: bias_noise_curve}(a), the mean error caused by duration bias is much larger than that of noisy watching in KuaiRand and WeChat is a decreasing curve with the growth of duration. Moreover, the duration bias caused error in WeChat is smaller than KuaiRand in a large margin. In Fig.~\ref{fig: bias_noise_curve}(b), the mean error caused by noisy watching in WeChat is a increasing curve while that in KuaiRand is a flatten curve which is close to zero. Meanwhile, the noisy watching caused error in WeChat is greater than KuaiRand. In summary, the KuaiRand can be regard as a dataset mainly affected by duration bias while the WeChat can be regard as a dataset affected by both duratin bias and noisy watching. For WeChat, the duration bias dominate the error in short duration while noisy watching dominate the error in long duration.

Then we split both KuaiRand and WeChat into three equal frequency duration ranges and evaluate the performance of each method in the corresponding subset. The results are shown in Table~\ref{tab: group_result}. To better reveal the performance difference, we defined the improve percentage $Imp(\%)_m = \frac{V_m - V_{wt}}{V_{o} - V_{wt}}$ in each subset, where $V_m$ is the value of current method's performance, $V_{wt}$ is the value of Watch Time's performance and $V_{o}$ is the value of Oracle's performance. Actually, $Imp(\%)_m$ indicates the relative effect of debias and denoise in the current subset. 
% On KuaiRand, baselines achieves the performance very close to that of D$^2$Co(A) in the all duration range. As we discussed before, KuaiRand are mainly affected by duration bias, so the results on KuaiRand indicate that all these methods has similar debias performance. 
% \xujun{Despite the effect of only bias in KuaiRand? not clear!}, 
For KuaiRand, although it has only duration bias caused error, our method D$^2$Co(A) and D$^2$Co(S) still exceeds the baseline, which shows the superiority of our method not relying on the critical assumptions. On WeChat, baselines and D$^2$Co(A) have similar performance in short duration while D$^2$Co(A) outperform baselines significantly in long duration. Also note that for these debiased baselines, their performances in the long duration of WeChat (e.g., $(42,60$]) even showed declines relative to the Watch Time. As we discussed before, WeChat is affected by duration bias in short duration and noisy watching in long duration, so the results on WeChat indicate that our proposed D$^2$Co has the ability to mitigate the noisy watching, thus outperform other baselines in a large margin on the long duration videos of WeChat.

\begin{figure*}
    \includegraphics[width=1.0\textwidth]{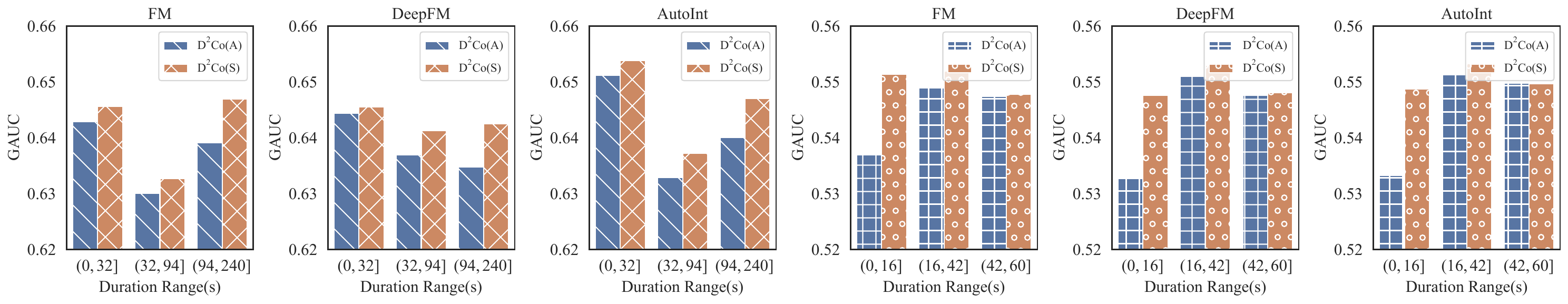}
    % \caption{The effect of sensitivity control in DeCo, w.r.t different backbone models. Top row: KuaiRand dataset; Bottom row: WeChat dataset.}
    \caption{The effect of sensitivity control in DeCo, w.r.t different backbone models. Left three: KuaiRand; Right three: WeChat.}
    \label{fig: sens_control}
\end{figure*}

\subsection{The Effectiveness of Sensitivity Control}
\label{subsec: effect of sense control}
In Theorem~\ref{thm: parameter sensitivity of D$^2$Co(A)}, we argue that the sensitivity of $w^{+}_{d}$ and $w^{-}_{d}$
produces different hazards for different datasets, and our sensitivity control correction reduces the sensitivity by controlling the corresponding sensitive parameters in different datasets. For KuaiRand, it has many records of the long watch time. These records make the sensitivity mainly dominated by $w^{+}_{d}$. For WeChat, it has many records of short watch time. These records make the sensitivity mainly dominated by $w^{-}_{d}$. Similarly, we divide the datasets into equal-frequency groups by duration range. The larger the duration, the longer the average watch time. Fig.~\ref{fig: sens_control} we present the GAUC of D$^2$Co(A) and D$^2$Co(S) in different duration ranges of two datasets. 
% In general, the performance improvement of D$^2$Co(S) with respect to D$^2$Co(A) is mainly in the long duration in KuaiRand and the short duration in WeChat. 
As we discussed, the bottleneck of KuaiRand is those long watch time records, so our proposed D$^2$Co(S) mainly outperforms D$^2$Co(A) in a large duration range (e.g., (94,240]). Meanwhile, the bottleneck of WeChat is those short watch time records, so our proposed D$^2$Co(S) mainly outperforms D$^2$Co(A) in a small duration range (e.g., (0,16]). In general, our proposed sensitivity-controlled correction is able to control the parameter sensitivity according to the bottleneck of different datasets, thus enhancing the original D$^2$Co(A).

% This is because there are more long watch time records in KuaiRand and short watch time records in WeChat, respectively. These records dominate the parameter sensitivity respectively, and our proposed sensitivity-controlled correction is able to control their sensitivity according to the bottleneck of different dataset, thus outperform the D$^2$Co(A).

% utilizing the standard affine correction to separate user interest suffers from the estimation error of bias and noise terms. Then we proposed sensitivity-controlled correction as a substitute of affine correction which has lower sensitivity. 

\subsection{The Effect of Hyper-Parameters}
\begin{figure*}
    \subfigure[FM]{
    \includegraphics[width=0.2\textwidth]{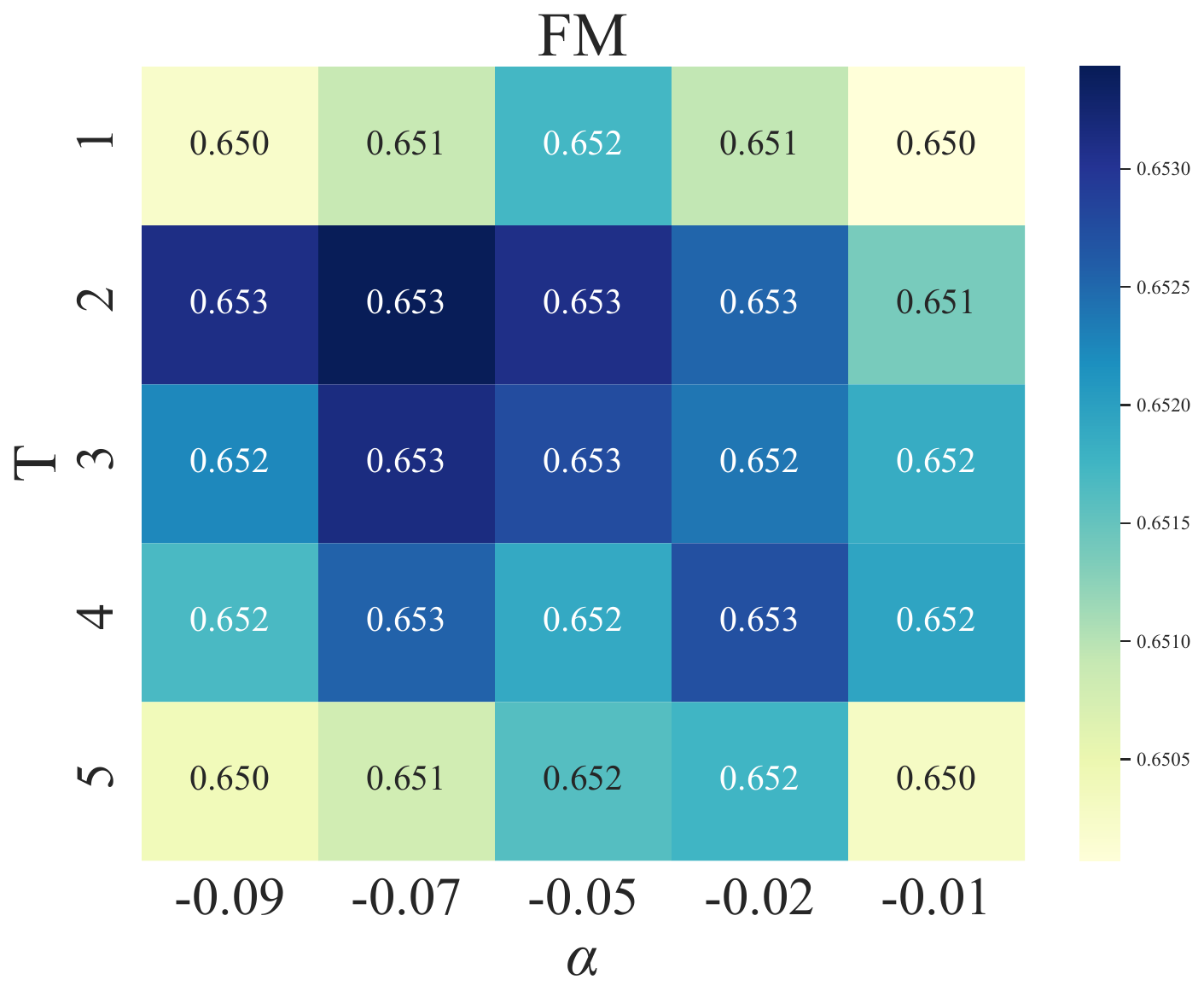}}
    \quad
    \subfigure[DeepFM]{
    \includegraphics[width=0.2\textwidth]{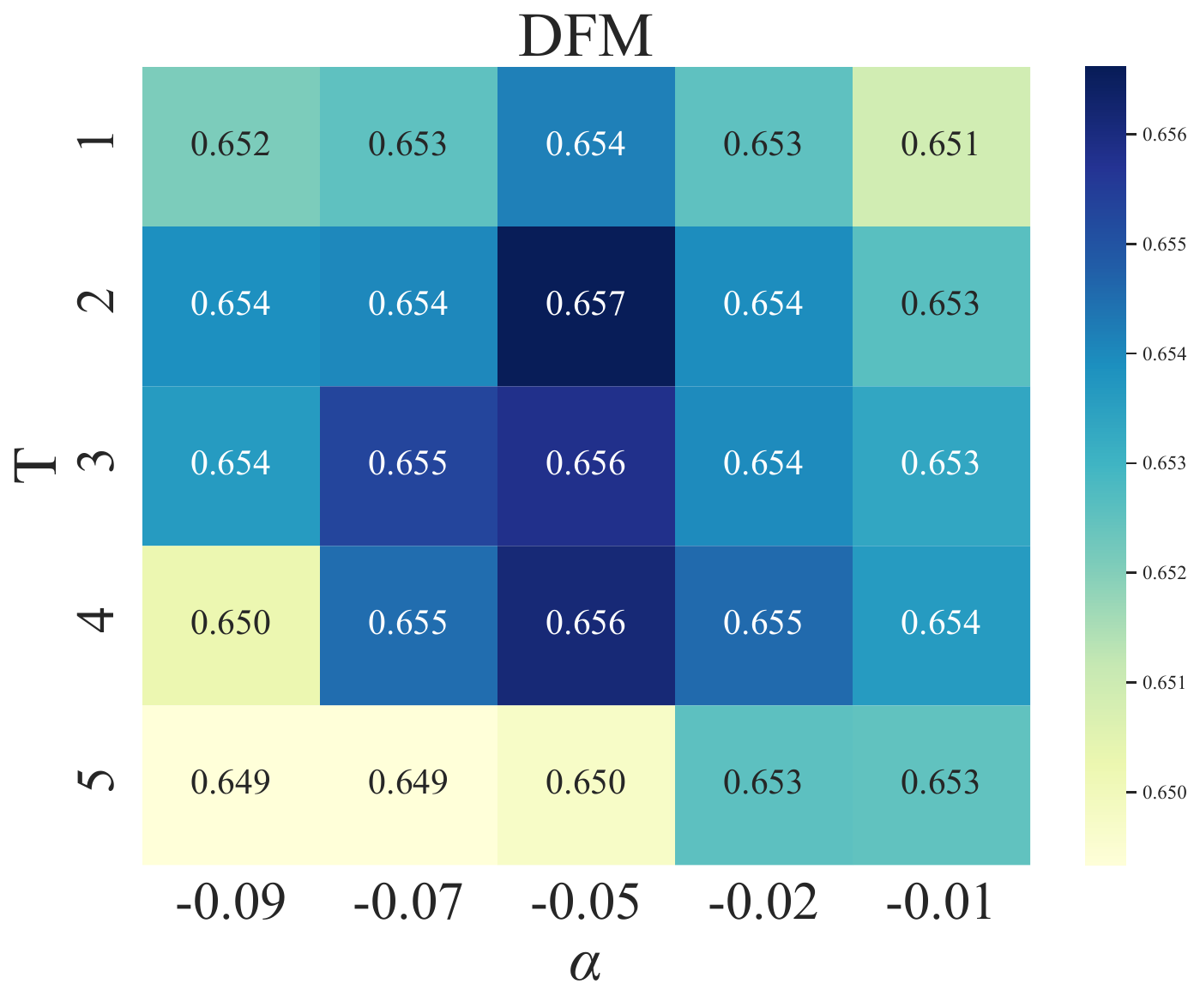}}
    \quad
    \subfigure[AutoInt]{
    \includegraphics[width=0.2\textwidth]{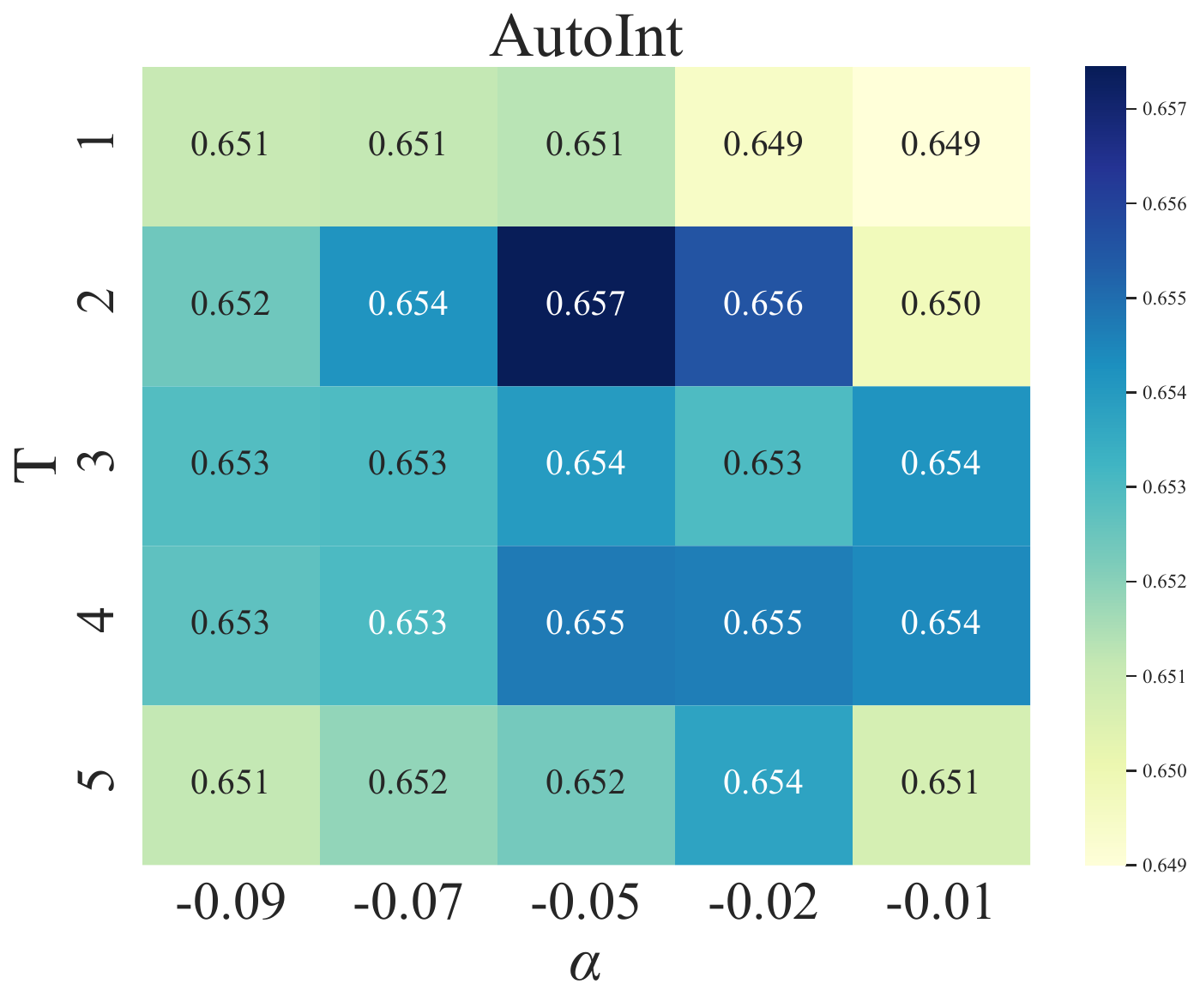}}
    \caption{Hyper-parameter sensitivity of D$^2$Co(S) w.r.t. different backbones in KuaiRand. Each cell denotes the corresponding GAUC.}
    \label{fig: hyper_param_sense}
\end{figure*}

There are two hyper-parameters of our proposed D$^2$Co. One is the size of the windows $T$ of frequency-weighted moving average in Eq.~\eqref{eq: moving_ave}. The larger the $T$, the smoother the bias and noise terms at adjacent times and the less specific the bias and noise terms themselves. The other is the sensitivity control term $\alpha$ of sensitivity-controlled correction in Eq~\eqref{eq: sense_control_D$^2$Co}. The larger the absolute value of $\alpha$, the greater the decrease in sensitivity of the corresponding bias and noise parameter, but the smaller the unbiasedness of estimated user interest. In most cases, $\alpha$ is set to a very small value. Both $T$ and $\alpha$ are essential for improving the performance of D$^2$Co. Fig.~\ref{fig: hyper_param_sense} illustrate the performance change of FM, DeepFM and AutoInt with different values of $T$ and $\alpha$. The figure indicates that different backbone recommendation models may have different reactions to the change of $T$ and $\alpha$. For FM (Fig.~\ref{fig: hyper_param_sense}(a)), the best hyper-parameter is $T\in\{2,3,4\} \land \alpha=-0.07$;For DeepFM (Fig.~\ref{fig: hyper_param_sense}(b)), the best hyper-parameter is $T\in\{2,3,4\} \land \alpha=-0.05$; For AutoInt(Fig.~\ref{fig: hyper_param_sense}(c)), the best hyper-parameter is $T=2 \land \alpha=-0.05$. In practice, it is necessary to adjust the hyper-parameters to make D$^2$Co perform best.

% \begin{table*}[]
% \caption{Relative improvement (\%) of D$^2$Co(S) to product baseline from online A/B testing}
% \resizebox{0.45\textwidth}{!}{
% \begin{tabular}{l|lllll}
% \hline
%                  & Impression  & VV & MWT      & PCR             & CTR             \\ \hline
% Day1             & 4.60\%            & 7.70\%            & 1.91\%          & 4.72\%          & 2.95\%          \\
% Day2             & 6.39\%            & 8.71\%            & 2.32\%          & 5.37\%          & 3.10\%          \\
% Day3             & 5.06\%            & 8.19\%            & 1.36\%          & 3.88\%          & 3.00\%          \\
% Day4             & 4.49\%            & 7.58\%            & -4.00\%         & 4.58\%          & 2.95\%          \\
% Day5             & 7.30\%            & 11.70\%           & 0.62\%          & 5.09\%          & 4.08\%          \\
% Day6             & 5.15\%            & 8.55\%            & -0.46\%         & 5.47\%          & 3.24\%          \\
% Day7             & 4.90\%            & 6.04\%            & 7.43\%          & 4.57\%          & 1.08\%          \\ \hline
% \textbf{Average} & \textbf{5.41\%}   & \textbf{8.35\%}   & \textbf{1.31\%} & \textbf{4.81\%} & \textbf{2.92\%} \\ \hline
% \end{tabular}}
% \label{tab: ab_test}
% \end{table*}

\begin{table*}[]
\caption{Relative improvement (\%) of D$^2$Co(S) to product baseline from online A/B testing}
\resizebox{0.6\textwidth}{!}{
\begin{tabular}{l|lllllll|l}
\hline
                  & Day1   & Day2   & Day3   & Day4    & Day5    & Day6    & Day7   & \textbf{Average} \\ \hline
Impression & 4.60\% & 6.39\% & 5.06\% & 4.49\%  & 7.30\%  & 5.15\%  & 4.90\% & \textbf{5.41\%}  \\
VV & 7.70\% & 8.71\% & 8.19\% & 7.58\%  & 11.70\% & 8.55\%  & 6.04\% & \textbf{8.35\%}  \\
MWT        & 1.91\% & 2.32\% & 1.36\% & -4.00\% & 0.62\%  & -0.46\% & 7.43\% & \textbf{1.31\%}  \\
PCR               & 4.72\% & 5.37\% & 3.88\% & 4.58\%  & 5.09\%  & 5.47\%  & 4.57\% & \textbf{4.81\%}  \\
CTR               & 2.95\% & 3.10\% & 3.00\% & 2.95\%  & 4.08\%  & 3.24\%  & 1.08\% & \textbf{2.92\%}  \\ \hline
\end{tabular}}
\label{tab: ab_test}
\end{table*}

\subsection{Online A/B Testing}
\label{sec: ab_test}
We conducted online A/B testing by deploying our D$^2$Co(S) in the video feeds of Huawei browser, a platform with tens of millions of daily active users (DAU), to evaluate its effectiveness in real video recommendation products. Specifically, we randomly split the users into the control and experimental groups. For the control group, the users were served by a highly-optimized deep CTR model without training by D$^2$Co(S). For the experimental group, the users were served by the same CTR model trained with D$^2$Co(S). Tabel~\ref{tab: ab_test} presents the relative improvements of the base model trained with D$^2$Co(S) on five online metrics: (1) Impression Volume; (2) Valid Viewing Volume (VV); (3) Mean Watch Time (MWT); (4) Play Complete Rate (PCR); (5) Click-Through Rate (CTR). The results show that the base model training with D$^2$Co(S) consistently outperforms the baseline by a large margin. One exception is the MWT, which fluctuates greatly in our A/B testing. The remarkable online improvements demonstrate the effectiveness of our proposed D$^2$Co in uncovering user interest from biased and noised watch time.
% \footnote{The Click here is not the actual click action but a binary indicator defined through watch time.}

% \input{6-result_discussion.tex}

\section{Conclusion}
In this study, we aim to discover user interest by watch time. Due to the effect of video duration, the watch time suffers from duration bias and noisy watching simultaneously. Current methods can only address duration bias while overlooking the noisy watching. Moreover, they rely on some critical assumptions to uncover the user interest, which may not hold in practice. To this end, we propose D$^2$Co to mitigate both duration bias and noisy watching. Specifically, we first employ a duration-wise Gaussian Mixture Model plus frequency-weighted moving average for estimating the bias and noise terms; then, we utilize a sensitivity-controlled correction function to separate the user interest from the watch time. The experiments on two public video recommendation datasets and online A/B testing indicate the effectiveness of the proposed D$^2$Co. 
\appendix
\section{\mbox{Detailed Analysis of Current Methods}}
This section shows the detailed analysis of the assumptions for the methods in Section~\ref{subsec: baseline analysis}.
% \subsection{Proof of Theorem 1}
% \label{proof of error}
% \begin{proof}
% \[
% \begin{split}
%     \left |\frac{w}{w_{\mathrm{max}}} - p^{r}_{\mathbf{x}}\right | &= \left | \frac{p^{r}_{\mathbf{x}}(w^{+}_{d} - w_{\mathrm{max}}) + (1 - p^{r}_{\mathbf{x}})w^{-}_{d}}{w_{\mathrm{max}}}\right | \\
%     &=\frac{w_{\mathrm{max}}-w^{+}_{d}}{w_{\mathrm{max}}}p^{r}_{\mathbf{x}} + \frac{w^{-}_{d}}{w_{\mathrm{max}}}(1-p^{r}_{\mathbf{x}})
% \end{split}
% \]
% \end{proof}

\subsection{The assumption of PCR}
\label{proof of prop pcr}
We can further rewrite PCR as:
\[
r^{\mathrm{PCR}}_{\mathbf{x}} = \frac{w}{d} = 
% \frac{w^{+}_{d}}{d}p^{r}_{\mathbf{x}} + \frac{w^{-}_{d}}{d}(1-p^{r}_{\mathbf{x}}) 
\frac{w^{+}_{d}p^{r}_{\mathbf{x}}+w^{-}_{d}(1-p^{r}_{\mathbf{x}})}{d}
= (\frac{w^{+}_{d}}{d} - \frac{w^{-}_{d}}{d})p^{r}_{\mathbf{x}} + \frac{w^{-}_{d}}{d}.
\]  
Then we have:
\[
\begin{split}
    &\forall i,j \in N, \quad (\frac{w^{+}_{d_i}}{d_i} - \frac{w^{-}_{d_i}}{d_i})p^{r}_{\mathbf{x_i}} + \frac{w^{-}_{d_i}}{d_i} > (\frac{w^{+}_{d_j}}{d_j} - \frac{w^{-}_{d_j}}{d_j})p^{r}_{\mathbf{x_j}} + \frac{w^{-}_{d_j}}{d_j}\Rightarrow p^{r}_{\mathbf{x_i}}>p^{r}_{\mathbf{x_j}} \\
    &iff.~ ~ \frac{w^{+}_{d_i}}{d_i} = \frac{w^{+}_{d_j}}{d_j} = C_1 \land \frac{w^{-}_{d_i}}{d_i} = \frac{w^{-}_{d_j}}{d_j} = C_2
\end{split}
\]
% \begin{proof}
% % Thus proving the proposition.
% \end{proof}

\subsection{The assumption of WTG}
\label{proof of prop wtg}
We can further rewrite WTG as:
\[
\begin{split}
    r^{\mathrm{WTG}}_{\mathbf{x}} &= \frac{w - \mu_{w}(d)}{\sigma_{w}(d)} 
    = \frac{(w^{+}_{d} - w^{-}_{d})p^{r}_{\mathbf{x}}+w^{-}_{d} - (w^{+}_{d} - w^{-}_{d})\mu_{p}(d) - w^{-}_{d}}{(w^{+}_{d} - w^{-}_{d})\sigma_{p}(d)} \\&
    = \frac{p^{r}_{\mathbf{x}} - \mu_{p}(d)}{\sigma_{p}(d)},
\end{split}
\]
where $\mu_{p}(d)$ and $\sigma_{p}(d)$ are the mean user interest and standard deviation of user interest in the video group with duration $d$, respectively. Then we have:
\[
\begin{split}
    &\forall i,j \in N,\frac{p^{r}_{\mathbf{x}_i} - \mu_{p}(d_i)}{\sigma_{p}(d_i)} > \frac{p^{r}_{\mathbf{x}_j} - \mu_{p}(d_j)}{\sigma_{p}(d_j)}\Rightarrow p^{r}_{\mathbf{x_i}}>p^{r}_{\mathbf{x_j}}, \\ & iff.~ ~\mu_{p}(d_i) = \mu_{p}(d_j) \land \sigma_{p}(d_i) = \sigma_{p}(d_j)
\end{split}
\]
% \begin{proof}
% % Thus proving the proposition.
% \end{proof}

\subsection{The assumption of D2Q}
\label{proof of prop d2q}
We can further rewrite D2Q as:
\[
\begin{split}
    r^{\mathrm{D2Q}}_{\mathbf{x}} &= \frac{|\mathcal{D}| - M\pi_{m}(w)}{|\mathcal{D}|} = \frac{\sum_{k}^{\frac{|\mathcal{D}|}{M}}\mathbb{I}(w>w_k)}{|\mathcal{D}|} \\
    &= \frac{\sum_{k}^{\frac{|\mathcal{D}|}{M}}\mathbb{I}\left((w^{+}_{d} - w^{-}_{d})p^{r}_{\mathbf{x}}+w^{-}_{d}>(w^{+}_{d} - w^{-}_{d})p^{r}_{\mathbf{x}_k}+w^{-}_{d}\right)}{|\mathcal{D}|} \\
    &= \frac{\sum_{k}^{|\frac{\mathcal{D}|}{M}}\mathbb{I}(p^{r}_{\mathbf{x}}>p^{r}_{\mathbf{x}_k})}{|\mathcal{D}|} = \frac{|\mathcal{D}| - M\pi_{m}(p^{r}_{\mathbf{x}})}{|\mathcal{D}|}
\end{split}
\]
where $\pi_{m}(p^{r}_{\mathbf{x}})$ is the ranking function of user interest $p^{r}_{\mathbf{x}}$. Then we have:
\[
\begin{split}
    &\forall i,j \in N,\frac{|\mathcal{D}| - M\pi_{m(i)}(p^{r}_{\mathbf{x_i}})}{|\mathcal{D}|} > \frac{|\mathcal{D}| - M\pi_{m(j)}(p^{r}_{\mathbf{x_j}})}{|\mathcal{D}|}\Rightarrow p^{r}_{\mathbf{x_i}}>p^{r}_{\mathbf{x_j}}, \\ & iff.~ ~\pi_{m(i)}(\cdot) = \pi_{m(j)}(\cdot)
\end{split}
\]
where $\pi_{m(i)}(\cdot)$ and $\pi_{m(j)}(\cdot)$ are the ranking function corresponding to the groups to which sample $i$ and $j$ belong. 

% \begin{proof}
% % Thus proving the proposition.
% \end{proof}

% \subsection{Proof of Theorem 2}
% \label{proof of sensitivity}
% \begin{proof}
% With the same small disturbance $\delta$, we have:
% \[
% \begin{split}
%     \mathbb{S}(w^{+}_{d}) &= \left|\frac{w-w^{-}_{d}}{w^{+}_{d}-w^{-}_{d}} - \frac{w-w^{-}_{d}}{w^{+}_{d} + \delta - w^{-}_{d}}\right| = \frac{\delta}{w^{+}_{d}-w^{-}_{d}}\left(\frac{w-w^{-}_{d}}{w^{+}_{d}-w^{-}_{d}+\delta}\right) \\
%     \mathbb{S}(w^{-}_{d}) &= \left|\frac{w-w^{-}_{d}}{w^{+}_{d}-w^{-}_{d}} - \frac{w - w^{-}_{d} - \delta}{w^{+}_{d} - w^{-}_{d} - \delta}\right| = \frac{\delta}{w^{+}_{d}-w^{-}_{d}}\left(\frac{w^{+}_{d}-w}{w^{+}_{d}-w^{-}_{d}-\delta}\right)
% \end{split}
% \]
% Since $\delta$ is a small disturbance, we have $w^{+}_{d}-w^{-}_{d}+\delta \approx w^{+}_{d}-w^{-}_{d}-\delta$. Then we only need to discuss the relationship between $w-w^{-}_{d}$ and $w^{+}_{d}-w$. Since,$w\in[w^{-}_{d},w^{+}_{d}]$, we have:
% \[
% \begin{split}
%     \mathbb{S}(w^{+}_{d}) &> \mathbb{S}(w^{-}_{d}),\quad \mathrm{if} ~ ~ w\in(\frac{w^{+}_{d}+w^{-}_{d}}{2}, w^{+}_{d}]\\
%     \mathbb{S}(w^{+}_{d}) &\leq \mathbb{S}(w^{-}_{d}),\quad \mathrm{if} ~ ~ w\in[w^{-}_{d},\frac{w^{+}_{d}+w^{-}_{d}}{2}]
% \end{split}
% \]
% \end{proof}

% \subsection{Proof of Proposition 4}
% \label{proof of lower sensitivity}

\begin{acks}
This work was funded by the National Key R\&D Program of China (2019YFE0198200), Beijing Outstanding Young Scientist Program NO. BJJWZYJH012019100020098, Intelligent Social Governance Interdisciplinary Platform, Major Innovation \& Planning Interdisciplinary Platform for the ``Double-First Class'' Initiative, Renmin University of China. Supported by fund for building world-class universities (disciplines) of Renmin University of China.
\end{acks}

\balance
\bibliographystyle{ACM-Reference-Format}
%\bibliography{sample-base}
%\bibliographystyle{plain}
\bibliography{ref}

\end{document}